\documentclass[10pt]{wlscirep}
\usepackage[utf8]{inputenc}
\usepackage[T1]{fontenc}

\usepackage{amsmath}
\usepackage{braket}
\usepackage{mathtools} 
\usepackage[justification=justified]{caption} 
\usepackage{lineno}
\usepackage{bm}
\usepackage{multirow}
\usepackage{threeparttable}

\title{Quantum teleportation mediated by surface plasmon polariton}

\author[1,$\dagger$]{Xin-He Jiang}
\author[1,$\dagger$]{Peng Chen}
\author[1,$\dagger$]{Kai-Yi Qian}
\author[1]{Zhao-Zhong Chen}
\author[1]{Shu-Qi Xu}
\author[1]{Yu-Bo Xie}
\author[1]{Shi-Ning Zhu}
\author[1,*]{Xiao-Song Ma}

\affil[1]{National Laboratory of Solid-state Microstructures, School of Physics,
Collaborative Innovation Center of Advanced Microstructures,
Nanjing University, Nanjing 210093, China}

\affil[*]{Xiaosong.Ma@nju.edu.cn}

\affil[$\dagger$]{These authors contributed equally to this work}

\begin{abstract}
Surface plasmon polaritons (SPPs) are collective excitations of free electrons propagating along a metal-dielectric interface. Although some basic quantum properties of SPPs, such as the preservation of entanglement, the wave-particle duality of a single plasmon, the quantum interference of two plasmons, and the verification of entanglement generation, have been shown, more advanced quantum information protocols have yet to be demonstrated with SPPs. Here, we experimentally realize quantum state teleportation between single photons and SPPs. To achieve this, we use polarization-entangled photon pairs, coherent photon-plasmon-photon conversion on a metallic subwavelength hole array, complete Bell-state measurements and an active feed-forward technique. The results of both quantum state and quantum process tomography confirm the quantum nature of the SPP mediated teleportation. An average state fidelity of 0.889$\pm$0.004 and a process fidelity of 0.820$\pm$0.005, which are well above the classical limit, are achieved. Our work shows that SPPs may be useful for realizing complex quantum protocols in a photonic-plasmonic hybrid quantum network.
\end{abstract}
\begin{document}

\flushbottom
\maketitle
%
%
\thispagestyle{empty}



\section*{Introduction}

The hybrid light-matter nature of surface plasmon polaritons (SPPs) allows light to be confined below the diffraction limit, opening up the possibility of subwavelength photonic device integration~\cite{Ozbay2006Science.311.189}. The quantum properties of SPPs originate from quantized surface plasma waves, and several quantum models have been proposed to describe the electromagnetic field of a plasmon~\cite{Ritchie1957PR.106.874, Hopfield1958PR.112.1555}. The quantization of SPPs has motivated many researchers to explore the fundamental quantum phenomena associated with them, for example, plasmon-assisted transmission of entangled photons~\cite{Altewischer2002, Fasel2005PRL.94.110501}, single-plasmon state generation and detection~\cite{Akimov2007Nature.450.7168, Heeres2013NNT.8.719}, quantum statistics and interference in plasmonic systems~\cite{Kolesov2009NP.5.470, Martino2012NL.12.2504, Martino2014PRApp.1.034004, Cai2014PRApp.2.014004, Fakonas2014NPh.8.317, Dheur2016SA.2.e1501574}, quantum logic operations~\cite{Wang2016NC.7.11490}, anti-coalescence of SPPs in the presence of losses~\cite{Vest2017Science.eaam9353} and quantum plasmonic N00N state for quantum sensing~\cite{Chen2018}. For reviews, see ref.~\cite{Tame2013QP, Li2017.66.144202}. Recently, some quantum properties of new plasmonic metamaterials have also been explored, such as coherent perfect absorption in plasmonic metamaterials with entangled photons~\cite{Altuzarra2017.4.2124}, testing hyper-complex quantum theories with negative refractive index metamaterials~\cite{Procopio2017NC.8.15044} and the active control of plasmonic metamaterials operating in the quantum regime~\cite{Uriri2018PRA.97.053810}.

These works motivate us to study and utilize the quantum properties of SPPs in more advanced quantum information protocols. Quantum teleportation uses entanglement as a resource to faithfully transfer unknown quantum states between distant nodes. Ever since it was first introduced by C.~H.~Bennett \textit{et al.}~\cite{Bennett1993PRL} and experimentally realized using photonic qubits~\cite{Bouwmeester1997, Boschi1998PRL}, quantum teleportation has become the essential protocol for establishing worldwide quantum networks~\cite{Cirac1997PRL.78.3221, Wehner2018Science.362.9288}. The teleportation distance has increased significantly over the last two decades~\cite{Marcikic2003Nature.421.509, Ursin2004Nature.430.849, Jin2010NP.4.376, Yin2012Nature.488.185, Ma2012Nature.489.269} and has recently been successfully extended to more than a thousand kilometres from the ground to a satellite~\cite{Ren2017Nature.549.70}. To build a quantum network with more functionalities, various physical systems are required with individual advantages in terms of transferring and processing the quantum state.

\section*{Results}

\subsection*{The conceptual scheme of SPP mediated quantum teleportation}

We experimentally realize the quantum state teleportation of a single photon to a single SPP, which is a single qubit consisting of collective electronic excitations typically involving {$\sim$}{$10^{6}$} electrons~\cite{Tame2013QP}. Our scheme is based on three qubits, which is first proposed by S.~Popescu~\cite{Popescu1995} and realized in experiment by D.~Boshi \textit{et al.}~\cite{Boschi1998PRL}. The conceptual framework of our experiment with the three-qubit scheme is shown in Fig.~\ref{fig1:ExpSet}(\textbf{a}). The entanglement between qubits 1 (Q1) and 2 (Q2), serving as the quantum channel, is generated from the entangled photon-pair source and distributed to Alice and Bob. An input state of qubit 0 (Q0) is sent to Alice. Alice performs a Bell-state measurement (BSM)~\cite{Boschi1998PRL}, projecting Q0 and Q1 randomly into one of the four Bell states, each with a probability of 25\%. Then, the outcomes of the BSM are sent to Bob through a classical communication (CC) channel. Q2 is sent to a subwavelength hole array sample patterned on a gold film at Bob's site to facilitate the photon-SPP-photon conversion~\cite{Ebbesen1998Nature.391.667}. There, the quantum state of Q2 is transferred to qubit 3 (Q3), carried by a single SPP. This SPP propagates along the surface of the sample and subsequently couples to an optical photon (Q4), which radiates towards detectors in the far field. According to the outcomes of the BSM, the corresponding unitary transformations (UTs) are applied to Q4. Finally, we perform quantum state tomography (QST)~\cite{Leonhardt1995PRL.74.4101, James2001PRA.64.052312} on Q4 and verify whether the quantum state teleportation from a single photon to a single SPP is successful by evaluating the quantum state fidelities of Q4 to Q0 and the quantum process fidelity of the whole procedure.

\subsection*{Subwavelength hole array and its characterization}
Figure~\ref{fig1:ExpSet}(\textbf{b}) shows a scanning electron microscopy (SEM) image of the subwavelength hole array used in our experiment. The gold film is perforated over a square area of 189$\times$189 $\mu$m$^2$ with periodic hole arrays by using a focused ion beam. The hole diameter and the period are 200 nm and 700 nm, respectively. The thickness of our metal film is 150-nm. Although the hole array reduces the direct photon transmission, it allows resonant excitation of the SPP~\cite{Ebbesen1998Nature.391.667}. 

The transmission spectrum of our sample is shown in Fig.~\ref{fig1:ExpSet}(\textbf{c}) and has a peak centred at approximately 809 nm with a full width at half maximum (FWHM) of $\sim$70 nm. The peak transmittance of the sample at 809 nm is approximately 0.8\%. The extraordinary optical transmission (EOT) observed in the subwavelength hole arrays is a typical resonant tunneling phenomenon which results from the constructive interference when the photons go through the holes~\cite{Ebbesen1998Nature.391.667,PhysRevLett.86.1114}. Compared with other works~\cite{Ebbesen1998Nature.391.667,Altewischer2002}, the total transmittance of our sample is slightly lower. The reason is that the transmission spectrum is very sensitive to the geometrical parameters of the system~\cite{PhysRevLett.92.183901,PhysRevLett.86.1114}. The imperfections during the fabrication can lead to the hole shape, period of the lattice as well as thickness and smoothness of the gold film departure from the nominal settings, thus resulting in the low transmission~\cite{altewischer2005sub}. Even setting the same parameters, the transmission of samples fabricated at different times has some obvious differences and is lower than 3\% due to the fabrication imperfections~\cite{Altewischer2002}. However, we only utilize the frequency information, i.e. peak position, instead of the transmittance in our teleportation experiment. Although our overall transmission is smaller than 2.5\%, it is still larger than the value predicted by the standard aperture theory~\cite{Ebbesen1998Nature.391.667}, which indicates that the EOT does happen in our sample. The transmission curves for different light polarizations are similar, indicating that our sample is nearly polarization-independent. The polarization insensitivity is due to the symmetry of the square lattice, as have been demonstrated in previous works~\cite{PhysRevLett.92.183901,Ren2007,ren2008polarization}. A numerical calculation based on the geometry of the array and the wavevector matching shows that this peak is associated with the ($\pm$1,$\pm$1) SPP modes at the glass-metal interface~\cite{Ghaemi1998PRB.58.6779}. These modes can excite the SPPs propagating along the four diagonal directions. We experimentally measure the SPP propagation with a laser and a charge-coupled device (CCD), as shown in Fig.~\ref{fig1:ExpSet}(\textbf{d}). By fitting to the SPP propagation along the diagonal direction, we estimate the $1/e$ decay length of the plasmonic mode to be $\sim$4.48$\pm$0.50 $\mu$m. See the Supplementary Information for more details on the numerical simulation, design of the hole array and characterizations of this device.

\subsection*{Realizing quantum teleportation between photon and SPP}

\begin{figure}[!htbp]
\centering
	\includegraphics[width=0.85\textwidth]{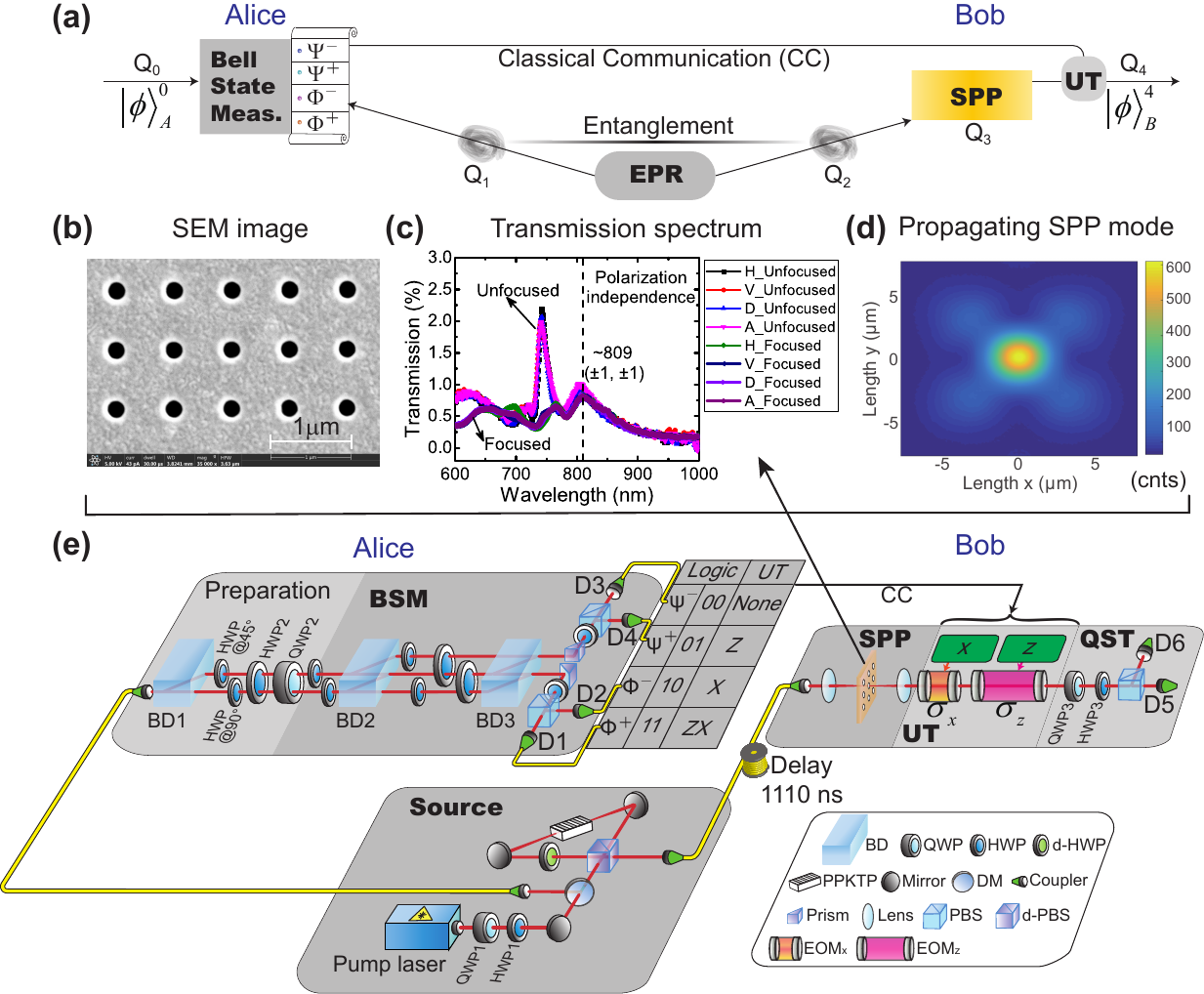}
	\caption{Experimental layout of the surface plasmon polariton (SPP) mediated quantum teleportation. (\textbf{a})~The conceptual framework of our experiment. At Alice's site, the input states are prepared using qubit 0 (Q0). An Einstein-Podolsky-Rosen (EPR) source generates two entangled qubits, Q1 and Q2. Q1 is sent to Alice for a Bell-state measurement (BSM)~\cite{Boschi1998PRL}. Q2 is sent to Bob to excite the SPP qubit, Q3. Through the photon-plasmon-photon conversion, the quantum states of the SPPs are transformed back to a photonic qubit, Q4. The outcomes of the BSM are sent to Bob using the classical communication (CC). Bob then applies a unitary transformation (UT) to Q4. As a result, the output state $\ket{\phi}^4_B$ is identical to $\ket{\phi}^0_A$; hence, teleportation is accomplished. (\textbf{b})~The SEM image of the subwavelength hole arrays with 200 nm diameter and 700 nm period. (\textbf{c})~Transmission spectrum of the hole arrays. The resonance at approximately 809 nm (dashed line) is the ($\pm$1,$\pm$1) mode, corresponding to the SPPs propagating along the diagonal direction. (\textbf{d})~The far-field image shows the SPP propagation mode. The units `counts' (cnts) is labelled below the colorbar. (\textbf{e})~Sketch of the experimental setup. The polarization-entangled source uses a type-II down-conversion Sagnac interferometer, where a $\chi^{(2)}$ nonlinear crystal (periodically poled KTiOPO$_4$, PPKTP) is coherently pumped by 405 nm laser light from clockwise and counter-clockwise directions. The central wavelength of the entangled signal (A) and idler (B) photons is approximately 810 nm. Photon A is sent to Alice. The polarization degree of freedom (DOF) (Q0) of photon A is used for preparing the six input states. The four Bell states are constructed using the path (Q1) and polarization (Q0) DOF of photon A. Photon B is sent to Bob. The polarization of photon B (Q2) is used to excite the SPPs. After undergoing a photon-plasmon-photon conversion, the quantum state of the SPPs (Q3) is transferred back to the photon (Q4). The results of the BSM (00, 01, 10, 11) are sent to Bob by CC and subsequently used to trigger the electro-optic modulators (EOMs, $\sigma_x$, $\sigma_z$) to apply the corresponding UTs ($I, \sigma_z, \sigma_x, i\sigma_y$).  The quantum state is finally analysed through quantum state tomography (QST). HWP: half-wave plate; QWP: quarter-wave plate; BD: beam displacer; DM: dichromatic mirror; d-PBS: dual-wavelength polarizing beam splitter.}
	\label{fig1:ExpSet}
\end{figure}

Figure~\ref{fig1:ExpSet}(\textbf{e}) presents a layout of our experimental setup. The entangled photon pairs are generated from spontaneous parametric down conversion, which is realized by embedding a periodically poled KTiOPO$_4$ (PPKTP) crystal in a Sagnac interferometer~\cite{Kim2006PRA.73.012316, Fedrizzi2007OE.15.15377}. The quantum state of photons A and B is similar to the singlet state:
\begin{align}\label{eq1:SourceState}
\ket{\Psi^{-}}_{AB} = \frac{1}{\sqrt{2}}(\ket{H}_{A}\ket{V}_{B}-\ket{V}_{A}\ket{H}_{B}),
\end{align}
which has a fidelity of approximately 98$\%$. $\ket{H}_{A}$ ($\ket{V}_{A}$) denotes the horizontal (vertical) polarization state of photon A. The same notation is used for photon B. We obtain coincidence counts at a rate of approximately 100 kHz with a pump power of 20 mW.

We employ the two-photon three-qubit scheme to realize the SPP mediated quantum teleportation~\cite{Boschi1998PRL, Jin2010NP.4.376}. The two-photon three-qubit scheme has the advantages that it avoids the very low detection rates caused by the simultaneous detection of three photons and allows a 100\% Bell state measurement~\cite{Popescu1995, Boschi1998PRL, Jin2010NP.4.376}. We note that two-photon scheme of teleportation has limitation as one can't use this scheme to teleport the quantum state of an independent photon which comes from outside. In our experiment, photons A and B are sent to Alice and Bob through single-mode fibre (SMF), respectively. We use photon A's polarization as Q0 and its path state as Q1. Photon B's polarization acts as Q2. First, we swap the entanglement between Q0 and Q2 (see equation~\eqref{eq1:SourceState}) to Q1 and Q2. We achieve this by sending photon A through a beam displacer (BD1 in Fig.~\ref{fig1:ExpSet}(\textbf{e})), which makes the horizontal polarized component undergo a lateral displacement into the left path mode (denoted as $\ket{l}$) and transmits the vertically polarized component directly (denoted as $\ket{r}$). The two-photon (A and B) three-qubit (Q0, Q1 and Q2) state can be written as
\begin{align}\label{eq2:WholeStateBD1}
	\ket{\Psi^{-}}^{012}_{AB} = \frac{1}{\sqrt{2}}(\ket{H}^{0}_{A}\ket{l}^{1}_{A}\ket{V}^{2}_{B}-\ket{V}^{0}_{A}\ket{r}^{1}_{A}\ket{H}^{2}_{B}).
\end{align}
Note that the superscripts are labelled for the qubit and the subscripts are labelled for the photon. Then, a 45$^\circ$-oriented HWP (HWP@45$^\circ$ in Fig.~\ref{fig1:ExpSet}(\textbf{e})) rotates the horizontal component ($\ket{H}_A$) to the vertical polarization ($\ket{V}_A$) in the left path, $\ket{l}$. Along the right path, $\ket{r}$, a 90$^\circ$-oriented HWP (HWP@90$^\circ$ in Fig.~\ref{fig1:ExpSet}(\textbf{e})) is used for phase compensation. After these two HWPs, the polarization state of photon A (qubit 0) is in $\ket{V}$ and is factorized out. The full state is as follows:
\begin{align}\label{eq2:WholeStateBD2}
	\ket{\Psi^{-}}^{012}_{AB} = \frac{1}{\sqrt{2}}\ket{V}^{0}_{A}\otimes(\ket{l}^{1}_{A}\ket{V}^{2}_{B}-\ket{r}^{1}_{A}\ket{H}^{2}_{B}).
\end{align}
Consequently, the initial entanglement between the polarization states of photons A and B is swapped into the path state of photon A (qubit 1) and the polarization state of photon B (qubit 2)~\cite{Giacomini2002PRA.66.030302, Takeda2013Nature.500.315}. 

The combination of HWP2 and QWP2 are then used to create the polarization state to be teleported (see Sec.~S5 of Supplementary Information), i.e.~$\ket{\phi}^{0}_{A}=\alpha\ket{H}^{0}_{A}+\beta\cdot \ket{V}^{0}_{A}$, where $\alpha$ and $\beta$ are two complex numbers satisfying $|\alpha|^2+|\beta|^2=1$. This process can be expressed as follows:
\begin{eqnarray}\label{eq3:InputState}
	\ket{\Psi^{-}}^{012}_{AB} &=& \left(\alpha\ket{H}^{0}_{A}+\beta\ket{V}^{0}_{A}\right)\otimes\frac{1}{\sqrt{2}}\left(\ket{l}^{1}_{A}\ket{V}^{2}_{B}-\ket{r}^{1}_{A}\ket{H}^{2}_{B}\right) \nonumber\\
	&=& \frac{1}{2}\left(i\sigma_y\ket{\phi}^{2}_{B}\ket{\Phi^{+}}^{01}_{A}+\sigma_x\ket{\phi}^{2}_{B}\ket{\Phi^{-}}^{01}_{A}-\sigma_z\ket{\phi}^{2}_{B}\ket{\Psi^{+}}^{01}_{A}+I\ket{\phi}^{2}_{B}\ket{\Psi^{-}}^{01}_{A}\right)
	\end{eqnarray}
Here the polarization (Q0) and path states (Q1) of photon A are used to construct the four Bell states: $\ket{\Psi^{\pm}}^{01}_A=\frac{1}{\sqrt{2}}(\ket{V}_{0}\ket{l}_{1}\pm\ket{H}_{0}\ket{r}_{1})$ and $\ket{\Phi^{\pm}}^{01}_A=\frac{1}{\sqrt{2}}(\ket{H}_{0}\ket{l}_{1}\pm\ket{V}_{0}\ket{r}_{1})$. Alice realizes a complete BSM using the polarization (Q0) and path (Q1) DOF of photon A with BD2 and BD3 (see Sec.~S5 of Supplementary Information for details). The outcomes of the BSM are sent from Alice to Bob via coaxial cables. 

Photon B (Q2) is delayed by a 222-m-long (corresponding to a temporal delay of $\sim$1110 ns) SMF and then sent to Bob. At Bob's site, Q2 is focused on the subwavelength hole arrays and converted to a single surface plasmon (Q3). As a result, we coherently transmit the quantum state of Q2 to Q3, which is carried by the single-mode collective electronic excitations of the SPP. Then, the SPP propagates along the surface of the sample and subsequently couples out to an optical photon (Q4), radiating into the far field. After the BSM is performed by Alice, the quantum state of Q4 is projected into a pure state and equals the input state $\ket{\phi}^{0}_{A}$ up to a local UT according to the BSM result (see equation~\eqref{eq3:InputState}). The local UTs are realized with two EOMs, which perform the required $\sigma_x$ and $\sigma_z$ operations. Collectively, the EOMs perform the $i\sigma_y$ operation. After these local UTs, the output state of Q4 is: $\ket{\phi}^{4}_{B}=\alpha\ket{H}^{4}_{B}+\beta\ket{V}^{4}_{B}$. Finally, we collect the photons into an SMF and perform QST on Q4.

\subsection*{The results of quantum state and process tomography}

\begin{figure}[!htbp]
\centering
	\includegraphics[width=0.6\textwidth]{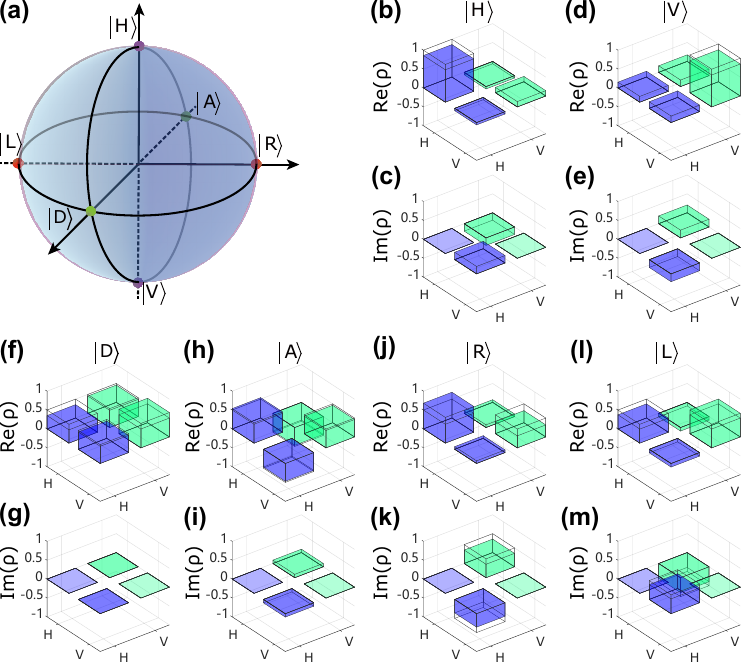}
	\caption{Reconstructed density matrices of the six teleported states. (\textbf{a})~The initial prepared states are $\ket{H}$, $\ket{V}$, $\ket{D}$, $\ket{A}$, $\ket{R}$, and $\ket{L}$ and are indicated by coloured dots on the Bloch sphere. (\textbf{b},~\textbf{d},~\textbf{f},~\textbf{h},~\textbf{j},~\textbf{l})~Real parts of the reconstructed density matrices for the six states. (\textbf{c},~\textbf{e},~\textbf{g},~\textbf{i},~\textbf{k},~\textbf{m})~Imaginary parts of the reconstructed density matrices for the six states. The ideal density matrix is shown as the wire grid. The representative data here are for experiments with a $\ket{\Phi^{+}}$ Bell-state measurement outcome with SPP. The reconstructed density matrices of the six states for all four Bell-state measurement outcomes are provided in the Supplementary Information.}
	\label{fig2:BlochDM}
\end{figure}

We prepare six input states of qubit 0: $\ket{H}$, $\ket{V}$, $\ket{D}$, $\ket{A}$, $\ket{R}$, and $\ket{L}$ (see Fig.~\ref{fig2:BlochDM}(\textbf{a})). Note that $\ket{D}=(\ket{H}+\ket{V})/\sqrt{2}$/$\ket{A}=(\ket{H}-\ket{V})/\sqrt{2}$, and $\ket{R}=(\ket{H}-i\ket{V})/\sqrt{2}$/$\ket{L}=(\ket{H}+i\ket{V})/\sqrt{2}$ stand for the diagonal/anti-diagonal linearly and right/left circularly polarized states of single photons, respectively.  

\begin{figure}[!htbp]
\centering
	\includegraphics[width=0.8\textwidth]{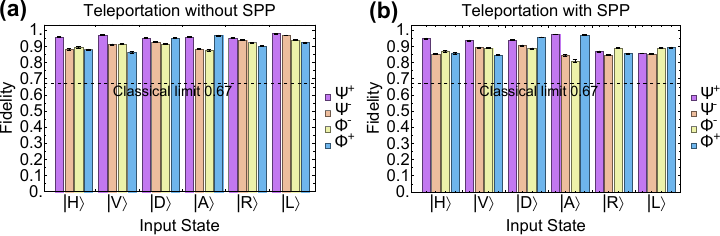}
	\caption{Quantum state fidelities of quantum teleportation for the six different input states: $\ket{H}$, $\ket{V}$, $\ket{D}$, $\ket{A}$, $\ket{R}$ and $\ket{L}$ with four Bell-state measurement results: $\ket{\Psi^{-}}$, $\ket{\Psi^{+}}$, $\ket{\Phi^{-}}$ and $\ket{\Phi^{+}}$. The different BSM outcomes are denoted with different colours. (\textbf{a})~The fidelities measured without the SPP involved.  We perform this measurement by moving the subwavelength hole array out from the setup. (\textbf{b})~The fidelities measured with the SPP involved. All the fidelities exceed the classical limit of 2/3 (dashed line). The error bars are calculated using a Monte Carlo routine assuming Poissonian statistics.}
	\label{fig3:Fid}
\end{figure}

To characterize the quantum teleportation mediated by the SPP, we perform single-qubit QST measurements on the teleported quantum states. In Fig.~\ref{fig2:BlochDM}(\textbf{b}-\textbf{m}), we show the real and imaginary parts of the reconstructed density matrices for different input states. With the reconstructed density matrices, we calculate the state fidelity $F=\prescript{}{ideal}{\left\langle\phi\right|}\rho\ket{\phi}_{ideal}$, where $\rho$ is the reconstructed density matrix and $\ket{\phi}_{ideal}$ is the ideal quantum state. The results of the quantum state fidelity after quantum teleportation are shown in Fig.~\ref{fig3:Fid}. For a comparison, we present the state fidelities both without and with photon-SPP-photon conversion. We can see from Fig.~\ref{fig3:Fid} that all the fidelities are well above the limit of $2/3$ that can be achieved using a classical strategy without employing entanglement~\cite{Massar1995PRL}. By averaging the single photon fidelities over all input states, we obtain an average fidelity of 92.67$\pm$0.32\% (without SPP) and 88.91$\pm$0.38\% (with SPP) for the retrieved initial states, including active feed-forward operations, which exceed the classical limit of 2/3 by more than 81-$\sigma$ and 58-$\sigma$ standard deviations~\cite{Massar1995PRL}. We note that the difference in the state fidelities between the cases without the SPP and with the SPP is mainly caused by: The excited SPP distorts the beam pattern and then leads to a lower contrast of the phase flip of the two EOMs. Quantitative analysis of the reduction in the achievable fidelity can be found in the Supplementary Information (Sec.~S7).

\begin{figure}[!htbp]
\centering
	\includegraphics[width=0.6\textwidth]{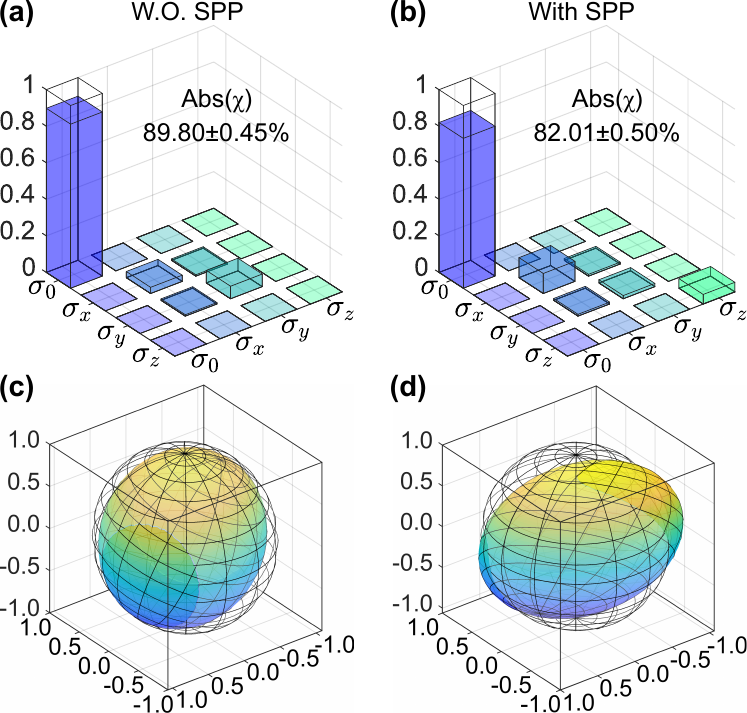}
	\caption{Results of quantum process tomography for the teleportation procedure. (\textbf{a})~The real part of the reconstructed process matrix $\chi$ without the SPP (W.O. SPP). The ideal process matrix has only one nonzero component $(\chi_{ideal})_{00}$=1, and we obtain a process fidelity of $F_{proc}=\text{Tr}(\chi_{ideal}\chi)=(89.80\pm0.45)\%$. (\textbf{b})~The real part of the reconstructed process matrix $\chi$ with the SPP (With SPP). The process fidelity is $F_{proc}=\text{Tr}(\chi_{ideal}\chi)=(82.01\pm0.50)\%$. (\textbf{c}, \textbf{d})~Bloch sphere representations of the process without (W.O.) (\textbf{c}) and with (\textbf{d}) the SPP involved. The plot shows how the input states lying on the surface of the initial Bloch sphere (meshed surface) are transformed by our teleportation protocol, with the output states lying on the solid surface.}
	\label{fig4:ProTomo}
\end{figure}
Since quantum teleportation is a quantum process, it is natural to quantitatively describe the whole process with quantum process tomography~\cite{Nielsen2010}. The reconstructed density matrices of the teleported quantum states allow us to fully characterize the teleportation procedure by quantum process tomography. We choose four input states ($\rho_{in} = \ket{H}\bra{H}, \ket{V}\bra{V}, \ket{D}\bra{D}, \ket{L}\bra{L}$) and their corresponding output states $\rho_{out}$ to benchmark the process of quantum teleportation. The effect of teleportation on $\rho_{in}$ is determined by the process matrix $\chi$, which is defined by $\rho_{out} = \sum_{l,k = 0}^3 \chi_{lk} \sigma_l \rho_{in} \sigma_k$, where $\sigma_{i}$ are the Pauli matrices with $\sigma_{0}$ being the identity operator. A perfect process matrix of quantum teleportation has only one nonzero component, $\chi_{00} = 1$, indicating that the input state is faithfully teleported without a reduction in the state fidelity. The real parts of the process matrix $\chi$ for the two situations (without and with the SPP) are shown in Fig.~\ref{fig4:ProTomo}(\textbf{a}) (without SPP) and Fig.~\ref{fig4:ProTomo}(\textbf{b}) (with SPP), respectively. The quantum process fidelities, i.e.~$\mathcal{F}_{proc}=\text{Tr}(\chi_{ideal}\chi)$, for our experiment without and with the SPP are 0.898$\pm$0.005 and 0.820$\pm$0.005, respectively. These fidelities correspond to 80-$\sigma$ and 64-$\sigma$ violations over the classical bound of 0.5~\cite{Nielsen2002PLA, Ma2012Nature.489.269}. A single-qubit quantum process, including quantum teleportation, can be represented graphically by a deformation of the Bloch sphere subjected to the quantum process~\cite{Nielsen2010}. As shown in Figs.~\ref{fig4:ProTomo}(\textbf{c}) (without SPP) and~\ref{fig4:ProTomo}(\textbf{d}) (with SPP), the ideal input states of Q0 are denoted as the states lying on the meshed surface of the Bloch sphere. After the photon-to-SPP quantum teleportation, the initial Bloch spheres are deformed into an anisotropic ellipsoids as shown in the solid blue-yellow colour, corresponding to the final output states. 

%

\section*{Discussion}

\noindent Note that the transmission losses reduce the coincidence count rate in our experiment. Therefore, we have to increase the integration time to obtain enough coincidence counts (see Table S7 in the Supplementary Information) for obtaining statistical significance. However, the advantages of plasmonic systems are that they are very suitable for making miniaturized quantum devices. Their sizes can be reduced such that the quantum logic operations can be finished below the propagation distances before plasmons are lost~\cite{Wang2016NC.7.11490}. Recently, the new techniques of material growth and structure design have helped in greatly minimize or mitigate the influence of losses on the plasmonic devices~\cite{Boriskina17,Baburin19,haffner2018low}. In some cases, the losses can provide new insights into quantum physics, such as the lossy beam splitter exhibiting fermionic anti-coalescence behavior using surface plasmons~\cite{Vest2017Science.eaam9353,Faccio1336}.

In our present experiment, the average teleportation fidelity for the retrieved initial states is smaller for the case with SPP than that without SPP. As have been explained in Sec.~S7 of the Supplementary Information, the reduction of fidelity is caused by the low contrast of feed-forward operations. With EOMs of higher extinction ratio, it is possible to improve the fidelity. In addition, the fabrication techniques can also be optimized to improve the quality of the sample and alleviate the deterioration of the light beam. We would like to elaborate on the motivation of our work from two different perspectives: 1, From a fundamental perspective: although quantum teleportation has been demonstrated with many different physical systems, to the best of our knowledge, it has never been implemented with the plasmonic system, a system consisting of 10$^6$ electrons. It would be interesting to see if the quantum teleportation could work between two systems with such dramatic particle number difference, namely one photon vs 10$^6$ electrons. 2, From the application perspective, plasmonic devices allow us to implement quantum operation with orders-of-magnitude smaller dimension~\cite{Wang2016NC.7.11490} comparing to SiO$_2$ or Si integrated devices~\cite{wang2019integrated}. However, to realize more complicated quantum information protocols, such as quantum teleportation of quantum gates~\cite{gao2010teleportation} or quantum teleportation based quantum computation~\cite{gottesman1999demonstrating} with plasmonic systems, we have to firstly experimentally verify the feasibility of quantum teleportation via plasmons. We view our work as the decisive step towards that goal, as the average state fidelity of the teleportation mediated by SPP we obtained exceeds the classical limit of 2/3 by more than 58-$\sigma$ standard deviations, which shows the plasmonic system is robust against the reduction of fidelity.

In summary, we demonstrate faithful teleportation of quantum states from one qubit of a single photon to another qubit of an SPP. The photon-to-SPP quantum teleportation is completely characterized by quantum state and process tomography. The fidelities of the six teleported states all exceed the classical limit with tens of standard deviations. The process fidelities also exceed the classical limit with tens of standard deviations. These results conclusively confirm the quantum nature of teleportation from arbitrary unknown quantum states of a single photon to a single SPP. Our work is a further step towards exploring the fascinating quantum behaviours of SPPs. The comprehensive utilization of the quantum properties of SPPs in more advanced protocols will promote the rapid development of future quantum information processing with quantum plasmonic devices.

%

\section*{Data Availability}

\noindent The data used in current study are available from the corresponding author upon reasonable request.


\section*{Acknowledgements}

\noindent The authors thank Tong Wu for help in designing the optical circuitry of the BSM. We would like to acknowledge support from the National Key Research and Development Program of China (2017YFA0303700, 2017YFA0303704), the National Natural Science Foundation of China (Grant No. 11690032, 11674170, 11621091), the Innovation Group of the National Natural Science Foundation of China (No. 3704), the Natural Science Foundation of Jiangsu Province (No. BK20170010), the Program for Innovative Talents and Entrepreneurs in Jiangsu and the Fundamental Research Funds for the Central Universities, and Leading-edge technology Program of Jiangsu Natural Science Foundation (No. BK20192001).

\section*{Author Contributions}

\noindent X.-H.~J., P.~C.~and K.-Y.~Q.~contributed equally to this work. X.-H.~J., P.~C., K.-Y.~Q.~and X.-S.~M.~designed and performed the experiment. X.-H.~J., K.-Y.~Q.~and S.-Q.~X.~fabricated and characterized the SPP samples with the help of Y.-B.~X.. P.~C.~and X.-H.~J.~designed and tested the BSM module. Z.-Z.~C.~designed and characterized the photon source. X.-H.~J.~and K.-Y.~Q.~set up the SPP module and carried out classical measurements for the sample. X.-H.~J.~analysed the data. X.-H.~J., P.~C., K.-Y.~Q.~and X.-S.~M.~wrote the manuscript with input from all authors. S.-N.~Z.~and X.-S.~M.~conceived the work and supervised the whole project.

\section*{Additional Information}

\noindent Supplementary information is available in the online version of the paper. Reprints and permission information is available online at www.nature.com/reprints. Correspondence and requests for materials should be addressed to X.-S.~M.

\section*{Competing Interests}

\noindent The authors declare no competing financial interests.

\clearpage
\setcounter{equation}{0}
\setcounter{figure}{0}
\setcounter{table}{0}
\makeatletter
\renewcommand{\theequation}{S\arabic{equation}}
\renewcommand{\figurename}{\textbf{Figure}}
\renewcommand{\thefigure}{\textbf{S\arabic{figure}}}
\renewcommand{\tablename}{\textbf{Table}}
\renewcommand{\thetable}{\textbf{S\arabic{table}}}
\renewcommand{\thesection}{\textbf{S\arabic{section}}}
\makeatother

\begin{center}
    \textbf{\LARGE\sffamily SUPPLEMENTARY INFORMATION: Quantum teleportation mediated by surface plasmon polariton}
\end{center}
\bigskip

%
%
%
%

\section{FDTD simulation and design of the sample}\label{sec:one}

\begin{figure}[!htbp]
\centering
	\includegraphics[width=0.7\textwidth]{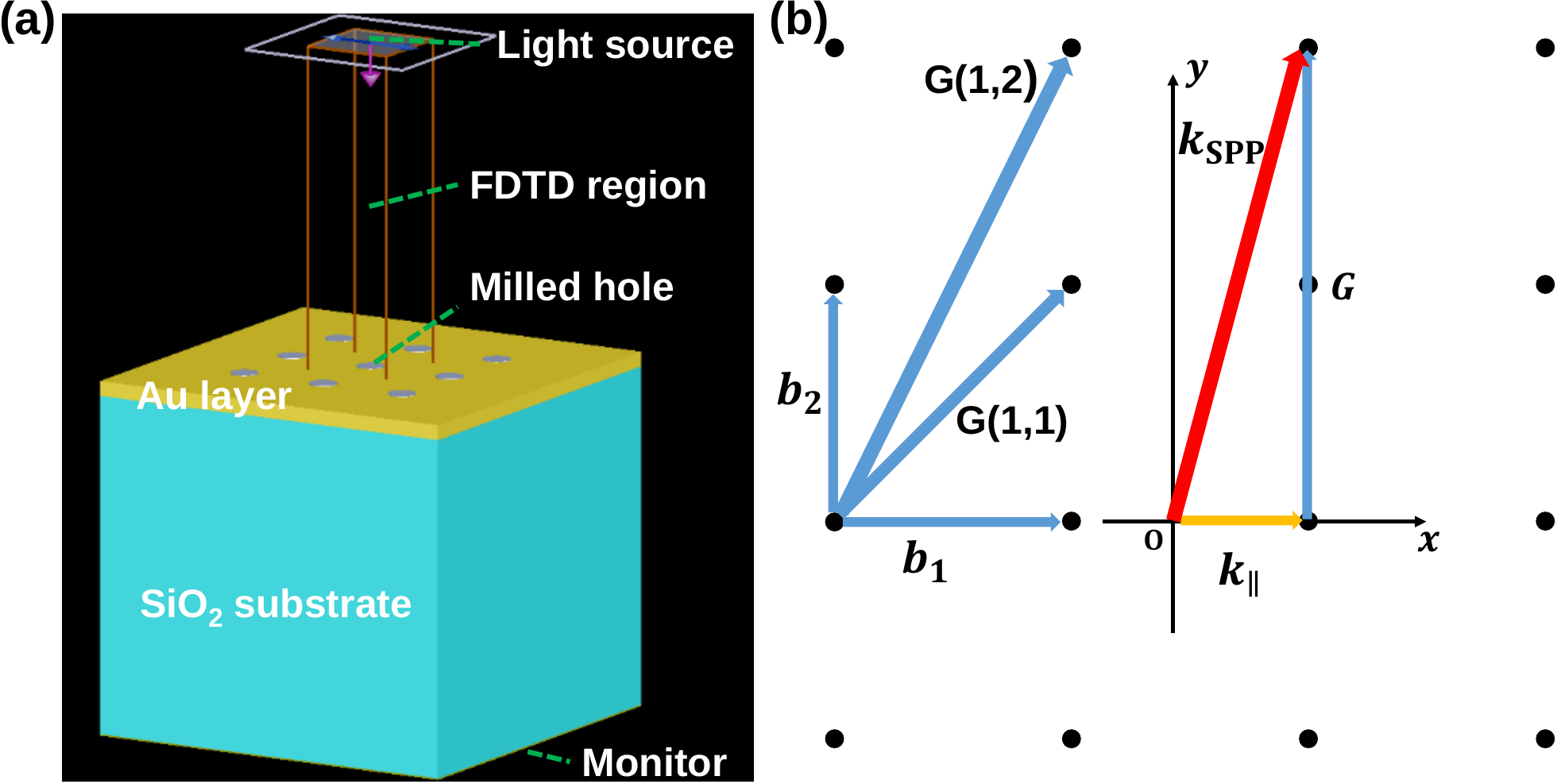}
	\caption{Structure of our simulation model and reciprocal lattice. (\textbf{a}) A mesh region of $p$ nm in $x$ and $y$ direction, and 10 $\mu$m in $z$ direction is used in the FDTD simulation. (\textbf{b}) Reciprocal lattice of hole array. The blue arrows indicate the reciprocal vectors. $\bm{k}_{\parallel}$ is the in-plane wave vector of the incident light. Excitation occurs when the wave vector of SPP $\bm{k}_\text{SPP}$ matches $\bm{k}_{\parallel}$ plus one reciprocal vector $\bm{G}$.}
	\label{SIfig:SimuStructure}
\end{figure}

To design the hole array with proper peak position and transmittance, we thoroughly perform finite-difference time-domain (FDTD) simulations with the commercial software Lumerical (FDTD solution, Lumerical Inc.). The mesh size of the primitive cell is 10 nm to ensure the accuracy of electromagnetic field calculations within the gold layer. The simulation model is shown in Fig.~\ref{SIfig:SimuStructure}.  The FDTD region is set to have the period $p$ in both $x$ and $y$ direction, and 10 $\mu$m in $z$ direction. The thickness of the gold layer is set to be $t$. The hole is perforated as a cylinder of diameter $d$. To provide a comprehensive guide to the fabrication, we simulate several periods (680, 690, 700, 710, 720 nm, etc.), thicknesses (130, 140, 150, 160, 170 nm, etc.) and diameters (180, 190, 200, 210, 220 nm, etc.). The boundary conditions in $x$ and $y$ directions are `periodical' and the $z$ direction uses the `Perfectly Matched Layer (PML)' boundary condition. The broadband light source is illuminated from the air-metal interface and the monitor placed at 5 $\mu$m away from the metal-glass interface is used to measure the transmitted light.

The dispersion relation for the resonant excitation of surface plasmons in a 2D lattice is 
\begin{equation}\label{eq19:DISPERSION}
\begin{aligned}
	\bm{k}_{\text{SPP}} &= \bm{k}_\parallel+\bm{G} \\
	\bm{G}(m_1,m_2)     &= m_1\bm{b}_1+m_2\bm{b}_2
\end{aligned}
\end{equation} 
where $\bm{k}_\parallel$ is the in-plane wave vector, $\bm{k}_{\text{SPP}}$ is the excited surface plasmon wave vector, $\bm{b}_1$ and $\bm{b}_2$ are the primitive vectors of reciprocal lattice (see Fig.~\ref{SIfig:SimuStructure}(b)), $|\bm{b}_1|=|\bm{b}_2|=2\pi/p$ with $p$ as the period of the hole array. This relation is guaranteed by the conservation of momentum. It is convenient to denote the wave vector of excited SPP $\bm{k}_\text{SPP}$ as the ($m_1$, $m_2$) mode. Generally, we can distinguish these modes according to the resonant wavelength of the excited SPP~\cite{SGhaemi1998PRB.58.6779}:
\begin{align}\label{eq16:Mode}
	\lambda(i, j)=\frac{p}{\sqrt{m_1^2+m_2^2}}\sqrt{\frac{\epsilon_{\text{S},\text{A}}\epsilon_{\text{M}}}{\epsilon_{\text{S},\text{A}}+\epsilon_{\text{M}}}}
\end{align}
where $m_1$ and $m_2$ are mode indices, $\epsilon_\text{M}$ is the dielectric constant of the metal, and $\epsilon_{\text{S},\text{A}}$ is the dielectric constant of substrate/air ($\epsilon_\text{S}$/$\epsilon_\text{A}$) in contact with the metal. For the sample used in our experiment, we can put the dielectric constants of gold (Au) and SiO$_2$/air into Eq.~\eqref{eq16:Mode} and obtain the mode corresponding to different resonant wavelengths. The calculated results indicate that the wavelength at approximately 810 nm is associated with the ($\pm1, \pm1$) mode that propagates along the four diagonal directions at the metal-substrate interface, as shown in Fig.~1(d) of main text. We adopt the optical dielectric constant of gold (real part $\varepsilon_1${}$\sim$-25.8 and imaginary part $\varepsilon_2${}$\sim$1.0-2.0) and dielectric constant of SiO$_2$ substrate (2.16-2.31) from the community~\cite{SOlmon2012PRB.86.235147,SGhaemi1998PRB.58.6779,SKRISHNAN20011}. The resonant wavelength is taken to be 810 nm. The period $p$ is taken as 700 nm. Substituting these parameters into Eq.~\eqref{eq16:Mode}, we calculate that the value of $m_1^2$+$m_2^2$ is approximately 1.90. Because $m_1$ and $m_2$ are integer mode index determining the matching reciprocal vector $\bm{G}$, the only possible mode labelling this resonant peak is ($m_1$=$\pm$1, $m_2$=$\pm$1). It should be pointed out that the transmission peak is not contributed by a single mode, the coherent admixture of different modes leads to the broadening and shift of the peak~\cite{SAltewischer2003JOSAB.20.1927}. For normal incident light and perfect square lattice, the ($1$, $1$) mode, ($1$, $-1$) mode, ($-1$, $1$) mode and ($-1$, $-1$) mode are degenerate. 

\begin{figure}[!htbp]
\centering
	\includegraphics[width=0.8\textwidth]{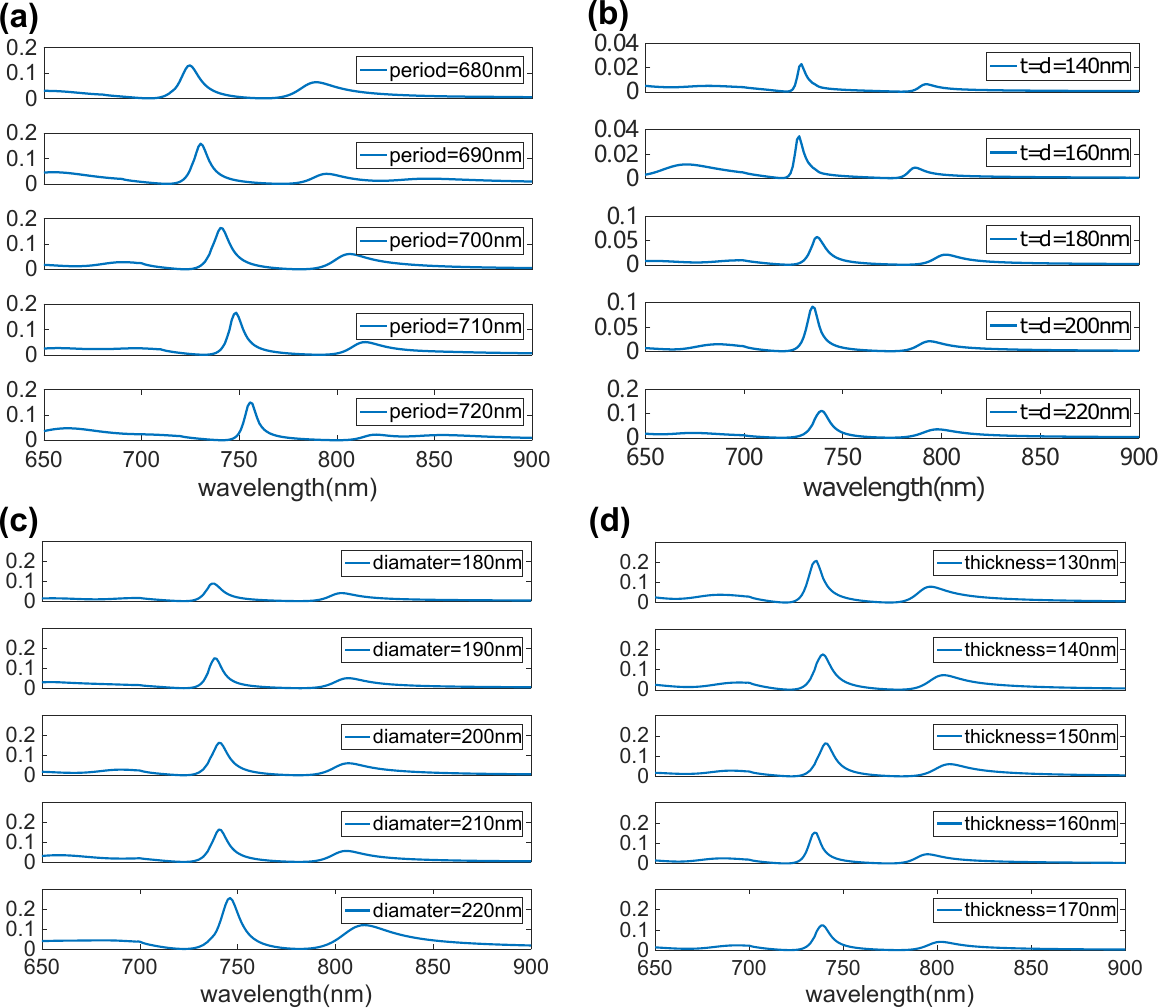}
	\caption{Simulation results for the influence of period, thickness $t$ and diameter $d$ on the transmission. (\textbf{a}), The influence of period on the peak position. We choose several periods around 700 nm. It shows an obvious red-shift when increasing the period. (\textbf{b}), (\textbf{c}), (\textbf{d}), The influence of thickness $t$ and diameter $d$ on the transmission spectra. In (\textbf{b}), we fix the ratio of $t/d$. In (\textbf{c}) and (\textbf{d}), we change the diameter and thickness. From the figure we can see that the peak position does not change so much as the ratio of $t$/$d$ remains a constant. The thickness and diameter have minor influences on the peak position, but obviously affect the transmittance.}
	\label{SIfig:SimuOfPAndTd}
\end{figure}

Figure~\ref{SIfig:SimuOfPAndTd} gives our simulation results of transmission spectrum using different periods $p$, diameters $d$ and thicknesses $t$. From Eq.~\eqref{eq16:Mode} we can see that the resonant wavelength increases when increasing the period of the hole array. This is confirmed by our FDTD simulations. The transmission peaks have a redshift as the period increases. It turns out to be an effective method to adjust the peak position by changing the period of the hole array. For the parameters of diameter and thickness, large thickness and small diameter will result in low transmittance and small full width at half maxima (FWHM). When the ratio $t/d$ is fixed, the peak position will not move too much. All the above analytical and numerical calculations give us a comprehensive guide for the design of the hole array. To reduce the photon losses in the single-photon experiment, we require large $d$, small $p$ and $t$ to increase the light transmittance. To obtain a peak around 810 nm, the period needs to be set at about 700 nm. To observe a clear ($\pm1, \pm1$) excitation mode, we require small $d$ and large $t$ to reduce the FWHM to prevent the mixture from other modes.

After theoretical calculations and simulations, we fabricate the sample based on the previous optimized parameters. We use quartz (SiO$_2$) as the substrate. Because of the poor adhesiveness of gold layer and SiO$_2$ substrate, a 3-nm-thick titanium bonding layer is deposited on the quartz. A 150-nm-thick gold layer is then deposited on the bonding layer using electron-beam evaporation. By means of focused ion beam (FIB), the hole array with a period of 700 nm and hole diameter of 200 nm is milled on the gold layer. The SEM image of our fabricated sample is shown in Fig.~1(b) of main text. Due to the finite image resolution and limited field of view (FOV), we have to move the sample stage to obtain a larger area of hole arrays. Nine small FOVs are put together closely to make a 3$\times$3 big array. Each FOV includes 90 periods and is therefore 63$\times$63 $\mu$m$^2$ in area. Consequently, the whole hole array has an area of approximately 189$\times$189 $\mu$m$^2$. We also fabricate other hole arrays with different periods and diameters in our experiment. The sample with 150-nm thickness, 700-nm period and 200-nm diameter has a small FWHM and clear ($\pm1, \pm1$) excitation mode. Moreover, it's transmission spectrum is polarization independent which guarantees that polarization-entangled states can be preserved in our single-photon experiment.

\section{Experimental characterization of the sample}\label{sec:two}

\begin{figure}[!htbp]
\centering
	\includegraphics[width=0.8\textwidth]{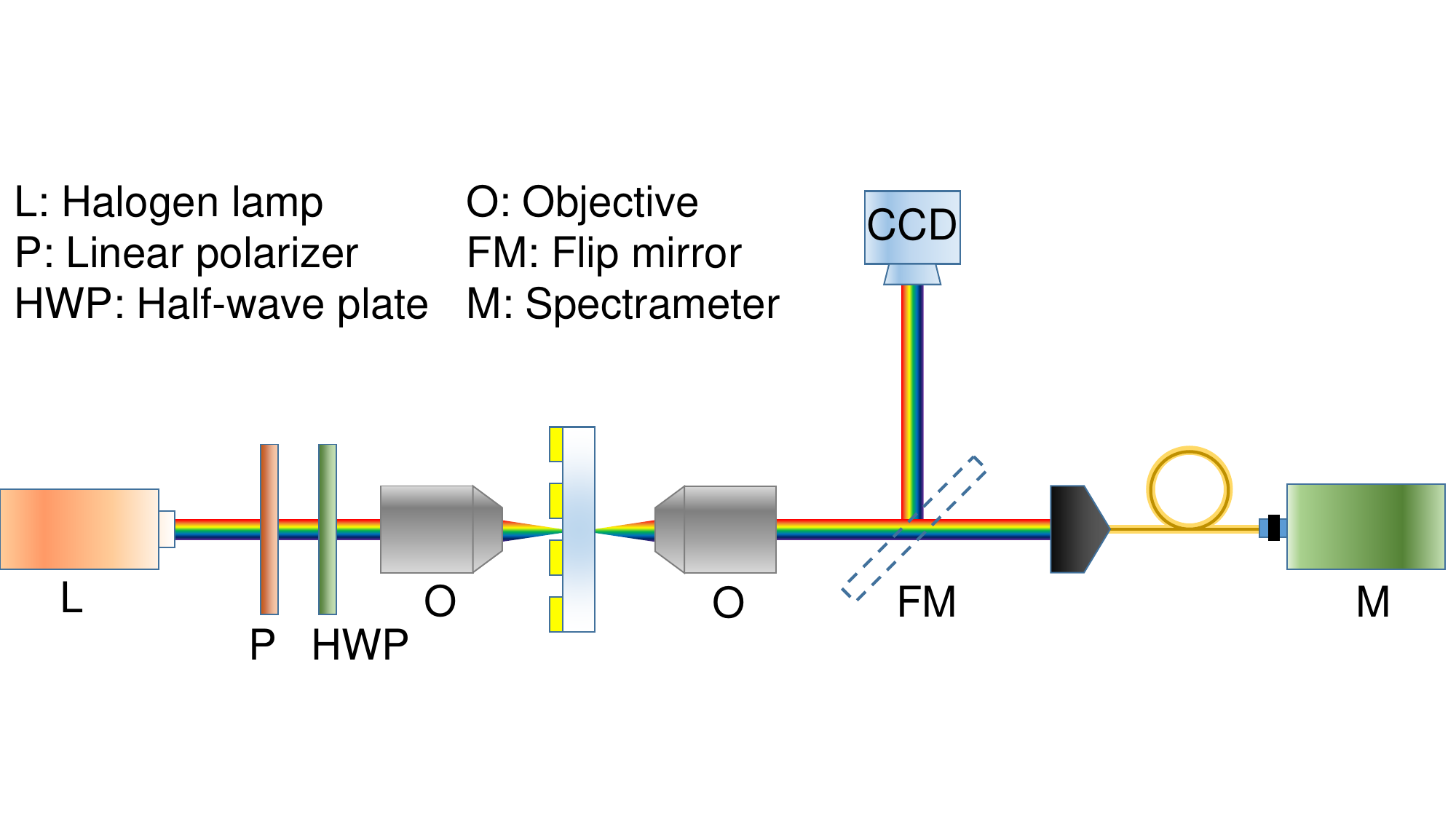}
	\caption{Experimental setup for transmission measurement and the measured and simulated results. The setup for transmission measurement. A white light source is focused on the sample and collected into the spectrometer. We also reflect the light into a charge-coupled device (CCD) for imaging and measuring the SPP excitation mode.}
	\label{SIfig:Charact}
\end{figure}
\begin{figure}[!htbp]
\centering
	\includegraphics[width=0.8\textwidth]{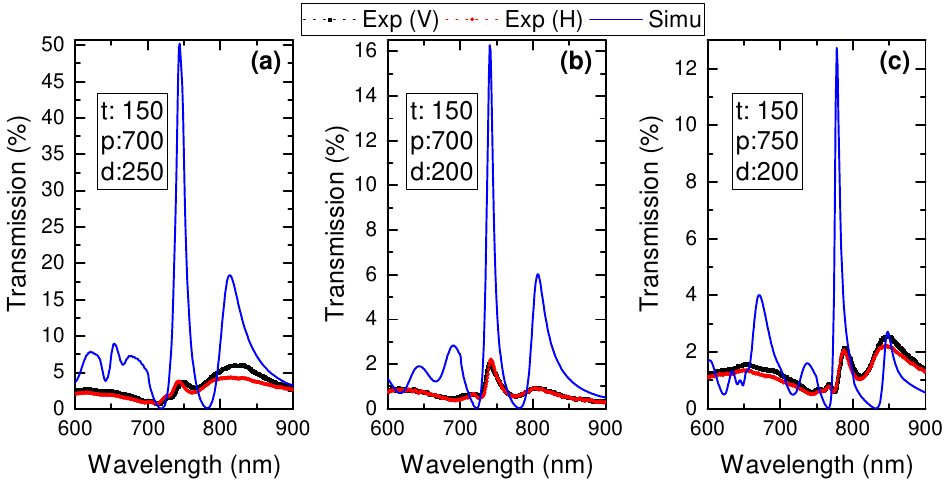}
	\caption{A comparison of the measured and simulated transmission spectra of three samples. (\textbf{a})~The transmission spectra for $t$=150, $p$=700 and $d$=250. (\textbf{b})~The transmission spectra for $t$=150, $p$=700 and $d$=200. (\textbf{c})~The transmission spectra for $t$=150, $p$=750 and $d$=200. Experimental transmission for different polarizations (V and H) are given. The measured peak position of transmission is in agreement with the simulation. The simulated transmittance is general larger than the experiment, which is caused by the imperfect fabrication of our sample. The difference between measurements and simulations are caused by the fabrication errors: the parameters by fabrication departure from the theoretical values and the imperfect junction caused by moving the sample stage.}
	\label{SIfig:Trans}
\end{figure}

The transmission spectra are measured to characterize the transmission properties of our sample. The experimental setup for the transmission measurement is shown in Fig.~\ref{SIfig:Charact}. We use a polarizer and a half-wave plate (HWP) after a halogen tungsten light source to get a linear polarized light. The sample is focused by a 40$\times$ objective. The transmitted light is collected by another 40$\times$ objective and collimated into the multimode fibre connected to a spectrometer. For comparison, we also measure the unfocused transmission spectra using a 20$\times$ magnification objective (the results are shown in Fig.~1(c) of main text). It is worth mentioning that the ellipticity of holes remarkably influences the transmission spectra of different polarizations. We adjust the focus of FIB and wait for a few minutes to release the stress to improve the circularity of the holes during the fabrication. Figure~\ref{SIfig:Trans} gives a comparison of the measurement and simulation results for three samples, i.e. $t$=150, $p$=700, $d$=250 (150700250), $t$=150, $p$=700, $d$=200 (150700200) and $t$=150, $p$=750, $d$=200 (150750200). From the figure we can see that the measured peak positions of the transmission spectra are consistent with those of simulations. The peak positions have a redshift when the period varies from 700 to 750 nm. When increasing the diameter of the holes from 200 to 250 nm, the transmittance has an obvious rise. We change the polarization of the incident light and find that the transmission spectra remain the same for the simulation results. For the experiment, the transmission spectra of different polarizations (V and H) have some differences due to the fabrication imperfection of the sample. The 150700200 hole array has an almost identical transmission spectrum for different polarizations and the peak position is around 810 nm, thus it is chosen for the teleportation experiment.

We also measure the SPP excitation mode at 25 different positions of the hole array and get an average figure with a charge-coupled device (CCD) (see Fig.~1(d) in the main text). As shown in Fig.~1(d) of the main text, the four lobes observed along the diagonal/anti-diagonal directions are in accord with the metal-substrate ($\pm1, \pm1$) mode. From the excitation mode, we obtain the SPP propagation distance to be approximately 4.48$\pm$0.50 $\mu$m ($1/e$ decay length along the diagonal direction, the error is obtained from the standard deviations of the 25 propagation distances). In addition, we also measure the coupling area in the SPP sample using a backward propagation light from the coupler. It turns out that an area with radius of 4.41$\pm$0.78 $\mu$m can be coupled into the single mode fibre (SMF). This demonstrates that the SPP genuinely participates in the teleportation process.

\section{Source of entangled photons}\label{sec:three}
We use a Sagnac interferometer~\cite{SKim2006PRA.73.012316, SFedrizzi2007OE.15.15377} to generate the entangled photons. A 20-mm long periodically-poled potassium titanyl phosphate (PPKTP) crystal is pumped by a 405 nm diode laser through type-II spontaneous parametric down-conversion (SPDC) process. The pump light generate orthogonally polarized photons with central wavelength of approximately 810 nm and a FWHM of approximately 0.5 nm. The visibility of polarization entanglement is typically $\sim$97\% (corresponding to a fidelity of 0.98). For the teleportation experiment, we use a pump power of typically 20 mW. The single photon count rate is approximately 0.5 MHz. The coincidence count rate is approximately 0.1 MHz. After one photon passes through the SPP sample, the single and coincidence counts will have a reduction due to the 0.8\% transmittance of the sample. After adding all other losses (coupling efficiency $\sim$30\%, propagation loss in fibre, EOM, objective, etc, $\sim$30\%), the total transmission efficiency of the SPP arm is approximately 0.1\%. The BSM arm has an efficiency of approximately 32\% (coupling efficiency $\sim$65\% and propagation loss $\sim$50\%). Therefore, single photon count rate and coincidence count rate with the SPP sample and BSM are approximately 1.5 kHz and 30 Hz (0.1 MHz $\times$ 0.1\% $\times$ 32\%), respectively.

\section{Certification of quantum properties of SPP}\label{sec:four}
In order to certify that the SPP can preserve the quantum correlation between the generated two photons, we perform the Bell-CHSH inequality tests~\cite{SBell1964Physics.1.195, SClauser1969PRL.23.880}. The violation of this inequality confirms the existence of entanglement between different particles in quantum systems. A hidden variable model requires that:
\begin{align}\label{eq1:CHSH}
	|S| = |E(\theta_1,\theta_2)-E(\theta_1,\theta_2')+E(\theta_1',\theta_2)+E(\theta_1',\theta_2')| \leqslant 2
\end{align}
where $\theta_1, \theta_2$ are angles of the measurement settings corresponding to photon A and B, respectively, and $E(\theta_1,\theta_2)$ is the correlation function with the settings ($\theta_1,\theta_2$). The correlation function is defined as
\begin{align}\label{eq2:CorrelaFun}
	E(\theta_1,\theta_2) = \frac{N_{++}-N_{+-}-N_{-+}+N_{--}}{N_{++}+N_{+-}+N_{-+}+N_{--}}
\end{align}
Here, $N_{mn}$ ($m,n=+,-$) represents the number of detected coincident events with the outcome $m$ for photon A and $n$ for photon B. The measurement settings for the two photons are $\left(\ket{\theta_1},\ket{\theta_1'}\right)$ and $\left(\ket{\theta_2},\ket{\theta_2'}\right)$, respectively. In order to maximally violate the Bell-CHSH inequality, we set the angles of measurement settings to be $\theta_1=0^\circ$, $\theta_1'=45^\circ$, $\theta_2=22.5^\circ$ and $\theta_2'=67.5^\circ$. The number of correlated events for each measurement base are listed in Table~\ref{TabCert4:BellIneq}. Our experiment gives a value of $S=2.551\pm$0.001 without the SPP and $S=2.281\pm$0.003 with the SPP, which are well above the classical bound 2. This indicates that the SPP can preserve the quantum correlation of the two photons.
\begin{table}[!htbp]
\centering
\caption{The measured coincidence counts $N_{mn}$ for the four base settings: $(\theta_1,\theta_2)$, $(\theta_1,\theta_2')$, $(\theta_1',\theta_2)$ and $(\theta_1',\theta_2')$. A total measurement time of five minutes per setting.}
\label{TabCert4:BellIneq}
\begin{tabular}{| l | c | c | c | c | c | c | c | c |}
	\hline
	\multirow{2}{*}{}     & \multicolumn{4}{c|}{Without SPP}          & \multicolumn{4}{c|}{With SPP}             \\
	\cline{2-9}
	                       & $N_{++}$ & $N_{+-}$ & $N_{-+}$ & $N_{--}$ & $N_{++}$ & $N_{+-}$ & $N_{-+}$ & $N_{--}$ \\
	\hline
	$(\theta_1,\theta_2)$  & 71100    & 5141100  & 6729000  & 275100   & 10800    & 842100   & 1107000  & 21600    \\
	\hline
	$(\theta_1,\theta_2')$ & 107400   & 1356300  & 1763700  & 1812900  & 22800    & 276600   & 364800   & 155700   \\
	\hline
	$(\theta_1',\theta_2)$ & 70200    & 4578600  & 5300100  & 413400   & 16800    & 664800   & 798600   & 54300    \\
	\hline
	$(\theta_1',\theta_2')$& 42000    & 6002400  & 5601000  & 320400   & 6300     & 975300   & 945300   & 22500    \\
	\hline
\end{tabular}
\end{table}

\section{Phase control of Bell-state analyser}\label{sec:five}
Bell-state measurement (BSM) plays a key role in a wide variety of quantum information processes such as entanglement swapping, quantum teleportation and quantum key distribution. Following the Rome scheme~\cite{SBoschi1998PRL, SJin2010NP.4.376}, we utilize both the path and polarization degree of freedoms of a single photon to encode the four Bell states to achieve the complete BSM. In our experiment, the four Bell states are defined as follows,
\begin{equation}\label{eqBellStates}
	\begin{aligned}
		\vert\Psi^\pm\rangle^{01}_{A} &= \frac{1}{\sqrt{2}}\left(\vert V \rangle^0_{A}\vert l \rangle^1_{A}\pm\vert H \rangle^0_{A}\vert r \rangle^1_{A}\right) \\
		\vert\Phi^\pm\rangle^{01}_{A} &= \frac{1}{\sqrt{2}}\left(\vert H \rangle^0_{A}\vert l \rangle^1_{A}\pm\vert V \rangle^0_{A}\vert r \rangle^1_{A}\right)
	\end{aligned}
\end{equation}

\begin{figure}[!htbp]
\centering
	\includegraphics[width=0.8\textwidth]{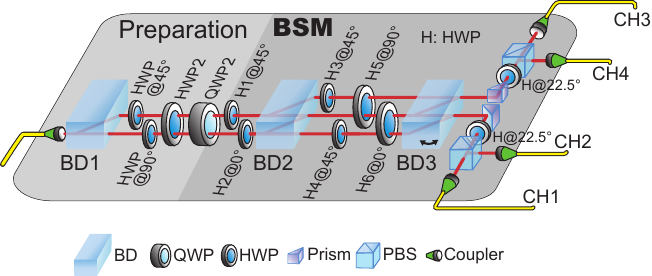}
	\caption{Experimental setup for state preparation and Bell-state measurement (BSM). The path-polarization entangled state $\vert\Psi^-\rangle^{012}_{AB}=\vert V \rangle^0_{A}\otimes\frac{1}{\sqrt{2}}(\vert l \rangle^0_{A}\vert V  \rangle^2_{B}-\vert r \rangle^1_{A}\vert H \rangle^2_{B})$  is generated after the photon A passes through the BD1, HWP@45$^\circ$ and  HWP@90$^\circ$. There are two wave plates, i.e.~HWP2 and QWP2, in both paths to prepare the states to be teleported. The whole setup behind QWP2 is used to realize the complete BSM.}
	\label{SIfig:BSM}
\end{figure}

After BSM, the state of photon A is projected to one of the four Bell states with equal probability and the state of photon B collapses to the unknown quantum state up to a unitary operation \{$i\sigma_y$, $\sigma_x$, $I$ and $\sigma_z$\}, where $\sigma_x$ , $\sigma_y$ and $\sigma_z$ are Pauli matrices. Hence, Alice needs to inform Bob about the outcomes of the BSM in real time via a classical communication channel. Then, Bob carries out the corresponding Pauli matrix operations to recover the original unknown quantum state according to the results of the BSM. The whole setup for the state preparation and BSM are shown in Fig.~\ref{SIfig:BSM}. All the components behind the quarter-wave plate 2 (QWP2) can be regarded as a black box to carry out the complete BSM. To make it clear, we give a detailed description of the BSM process with matrix representation.

The three path and two polarization modes allow us to construct a 6$\times$6 matrix representation for the transformation of each optical element. Correspondingly, the quantum state can be represented by a vector of six components. The unitary operation matrix of the BSM can be depicted by temporarily disregarding the accumulated phase from each element:
\begin{equation}
\mathbb{U}=\left(
\begin{array}{cccccc}
0&0&0&\frac{1}{\sqrt{2}}&-\frac{1}{\sqrt{2}}&0\\
0&0&0&-\frac{1}{\sqrt{2}}&-\frac{1}{\sqrt{2}}&0\\
0&0&-\frac{1}{\sqrt{2}}&0&0&-\frac{1}{\sqrt{2}}\\
0&0&\frac{1}{\sqrt{2}}&0&0&-\frac{1}{\sqrt{2}}\\
0&0&0&0&0&0\\
0&0&0&0&0&0\\
\end{array}\right)
\end{equation}
The four Bell states can be wrote respectively as the following column vectors:
\begin{equation*}
\vert\bm{\Phi}^+\rangle:\left(
\begin{array}{c}
0\\
0\\
1\\
0\\
0\\
1\\
\end{array}\right)
,
\vert\bm{\Phi}^-\rangle:\left(
\begin{array}{c}
0\\
0\\
1\\
0\\
0\\
-1\\
\end{array}\right)
,
\vert\bm{\Psi}^+\rangle:\left(
\begin{array}{c}
0\\
0\\
0\\
1\\
1\\
0\\
\end{array}\right)
,
\vert\bm{\Psi}^-\rangle:\left(
\begin{array}{c}
0\\
0\\
0\\
1\\
-1\\
0\\
\end{array}\right)
\end{equation*}
When the BSM is finished, the photon A is projected to different Bell states and comes out from individual port. Each port has corresponding vector as follows:
\begin{equation*}
\mathbf{CH1}:\left(
\begin{array}{c}
0\\
0\\
1\\
0\\
0\\
0\\
\end{array}\right)
,
\mathbf{CH2}:\left(
\begin{array}{c}
0\\
0\\
0\\
1\\
0\\
0\\
\end{array}\right)
,
\mathbf{CH3}:\left(
\begin{array}{c}
1\\
0\\
0\\
0\\
0\\
0\\
\end{array}\right)
,
\mathbf{CH4}:\left(
\begin{array}{c}
0\\
1\\
0\\
0\\
0\\
0\\
\end{array}\right)
\end{equation*}
It is not difficult to verify the correspondence of each port to each Bell state by means of matrix operations. In Table~\ref{TabBSM51:Port}, we give a detailed list for the mapping of these four ports. From above analysis, we can see that the four Bell states can be fully distinguished using our setup. Therefore, we can in principle perform deterministic teleportation.
\begin{table}[!htbp]
\centering
\caption{Exiting port correspondence for the four Bell states.}
\label{TabBSM51:Port}
\begin{tabular}{| l | c | c | c | c |}
	\hline
	Ports & $CH1$ & $CH2$ & $CH3$ & $CH4$ \\ \hline
	Bell states & $\vert\Phi^+\rangle$ & $\vert\Phi^-\rangle$ & $\vert\Psi^-\rangle$ & $\vert\Psi^+\rangle$ \\
	\hline
\end{tabular}
\end{table}

\begin{figure}[!htbp]
\centering
	\includegraphics[width=0.8\textwidth]{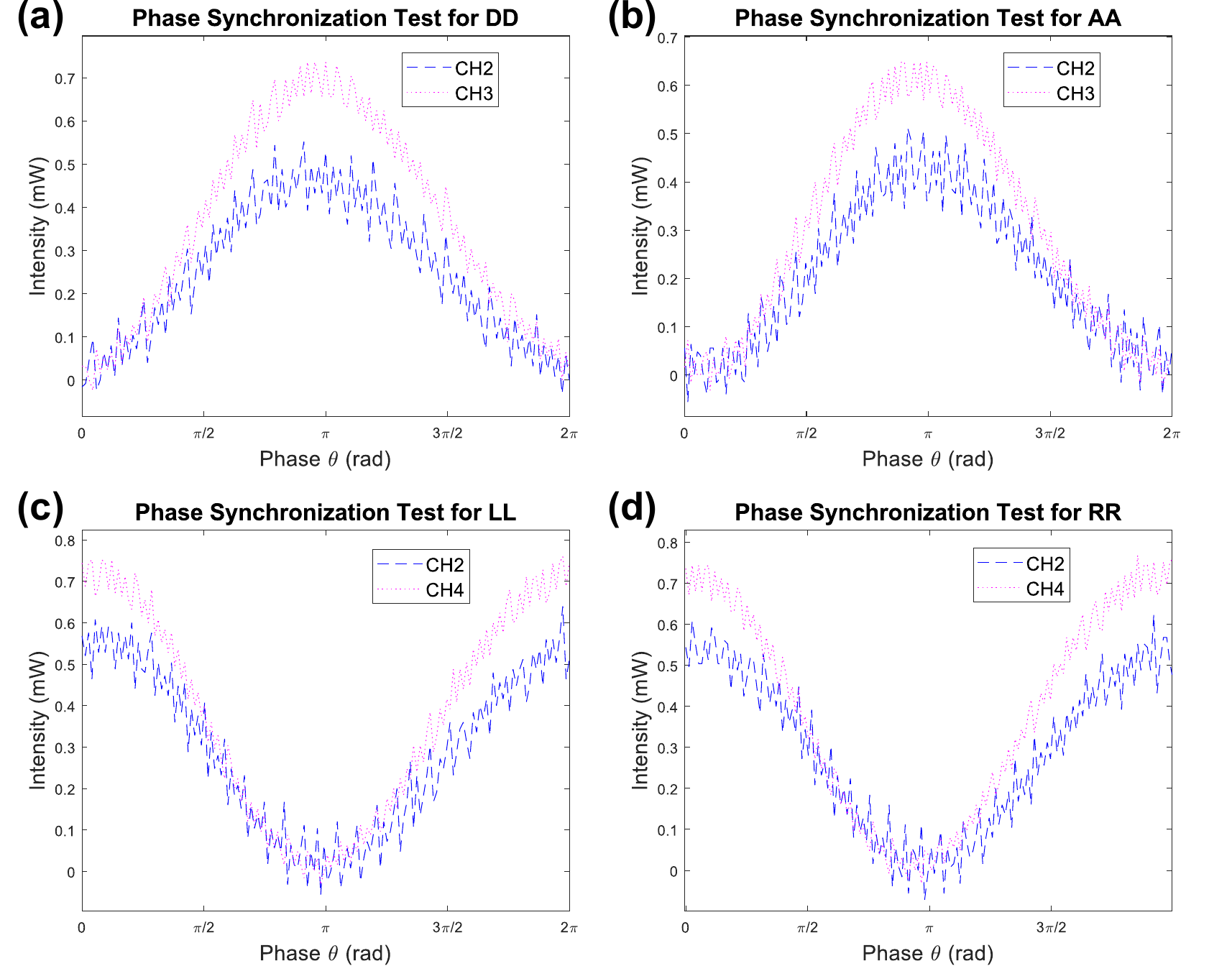}
	\caption{Phase synchronization test for (\textbf{a}) DD, (\textbf{b}) AA, (\textbf{c}) RR and (\textbf{d}) LL. The first letter denotes the polarization state of incident light and the second stands for the prepared state. In this case, the incident and prepared states are identical. Through optimization, CH2 is in sync with CH3 for DD and AA and in sync with CH4 for RR and LL. These four tests show the phase synchronization for different incident light polarizations when the polarization of prepared state is the same. D: Diagonal polarization; A: Antidiagonal polarization; R: Right circularly polarization; L: Left circularly polarization.}
	\label{SIfig:DD}
\end{figure}
\begin{figure}[!htbp]
\centering
	\includegraphics[width=0.8\textwidth]{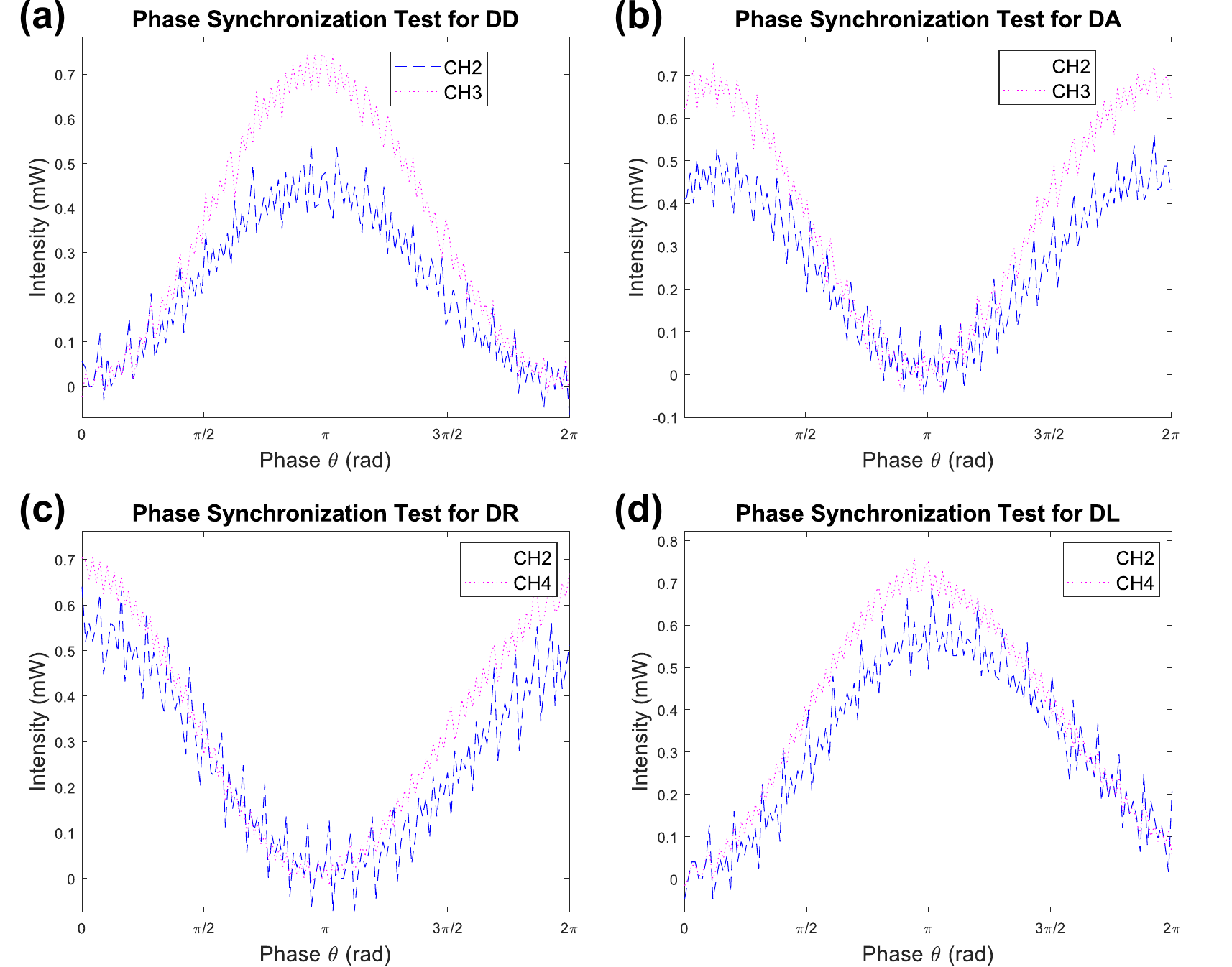}
	\caption{Phase synchronization test for (\textbf{a}) DD, (\textbf{b}) DA, (\textbf{c}) DR and (\textbf{d}) DL. The meanings of these letters are the same as in Fig.~\ref{SIfig:DD}. In this case, the incident state is always D and four prepared states are D, A, R and L, respectively. Note that (\textbf{a}) is the same as Fig.~\ref{SIfig:DD}(a). Through optimization, CH2 is in sync with CH3 for DD and DA and in sync with CH4 for DR and DL. As a complementary part of Fig.~\ref{SIfig:DD}, these tests can guarantee the perfect phase synchronization for the same incident light with general prepared states.}
	\label{SIfig:DA}
\end{figure}
The challenge for our BSM setup is the phase control which includes accumulated relative phase between different paths and phase synchronization of different Bell states $\vert\Psi^\pm\rangle$ and $\vert\Phi^\pm\rangle$. In our experiment, we address the issues by tuning the pitch angle of H3@45$^\circ$ or H4@45$^\circ$. Before the single photon experiment, classical light with different polarizations is used to calibrate the phase of the BSM interferometer. The interference visibility is optimized to be approximately 96\% with a piezoelectric ceramics attached on the BD3 which is driven by an external triangle wave signal. Furthermore, we make a series of phase synchronization test using different incident polarized light ($D$, $A$, $R$, $L$) with polarizer+HWP+QWP (not shown in Fig.~\ref{SIfig:BSM}) and prepared states ($D$, $A$, $R$, $L$) with HWP2+QWP2 (see Fig.~\ref{SIfig:BSM}). The results of phase synchronization for seven representative settings ($DD$, $DA$, $DL$, $DR$, $AA$, $RR$, $LL$, where the first letter denotes the polarization state of incident light and the second letter stands for the prepared state) are given in Figs.~\ref{SIfig:DD} and~\ref{SIfig:DA}. We only need to test the phase synchronization between port CH1 (CH2) and port CH3 (CH4) on account of the instinctive phase identity of CH1 (CH3) and CH2 (CH4) because they go through the same phase difference caused by BD2 and BD3 and exit from the same PBS. The light from the four Bell ports are coupled into SMF and detected with photodetector. By changing the voltage applied on the ceramics, we can scan the phase and obtain the variation of intensity with respect to the phase. From Figs.~\ref{SIfig:DD} and~\ref{SIfig:DA} we can see that the maximum and minimum almost appear at the same phase location, which indicates that the phase can be synchronized for all the Bell ports using our optimization methods. Note that the intensities of the four Bell ports may have different maximum because of the varying power of incident polarized light and the different coupling efficiencies. Figures~\ref{SIfig:DD} and~\ref{SIfig:DA} show that the four Bell ports have some symmetric phase relations with respect to each other for different combinations of incident and prepared polarized states. After making the synchronization of four Bell ports, we rotate the azimuth of BD3 to move the phase to the location of maximal contrast (at {$\sim$}{$\pi$} in Figs.~\ref{SIfig:DD} and~\ref{SIfig:DA}).

\section{Feed-forward unitary transformations}\label{sec:six}
\begin{figure}[!htbp]
\centering
	\includegraphics[width=0.8\textwidth]{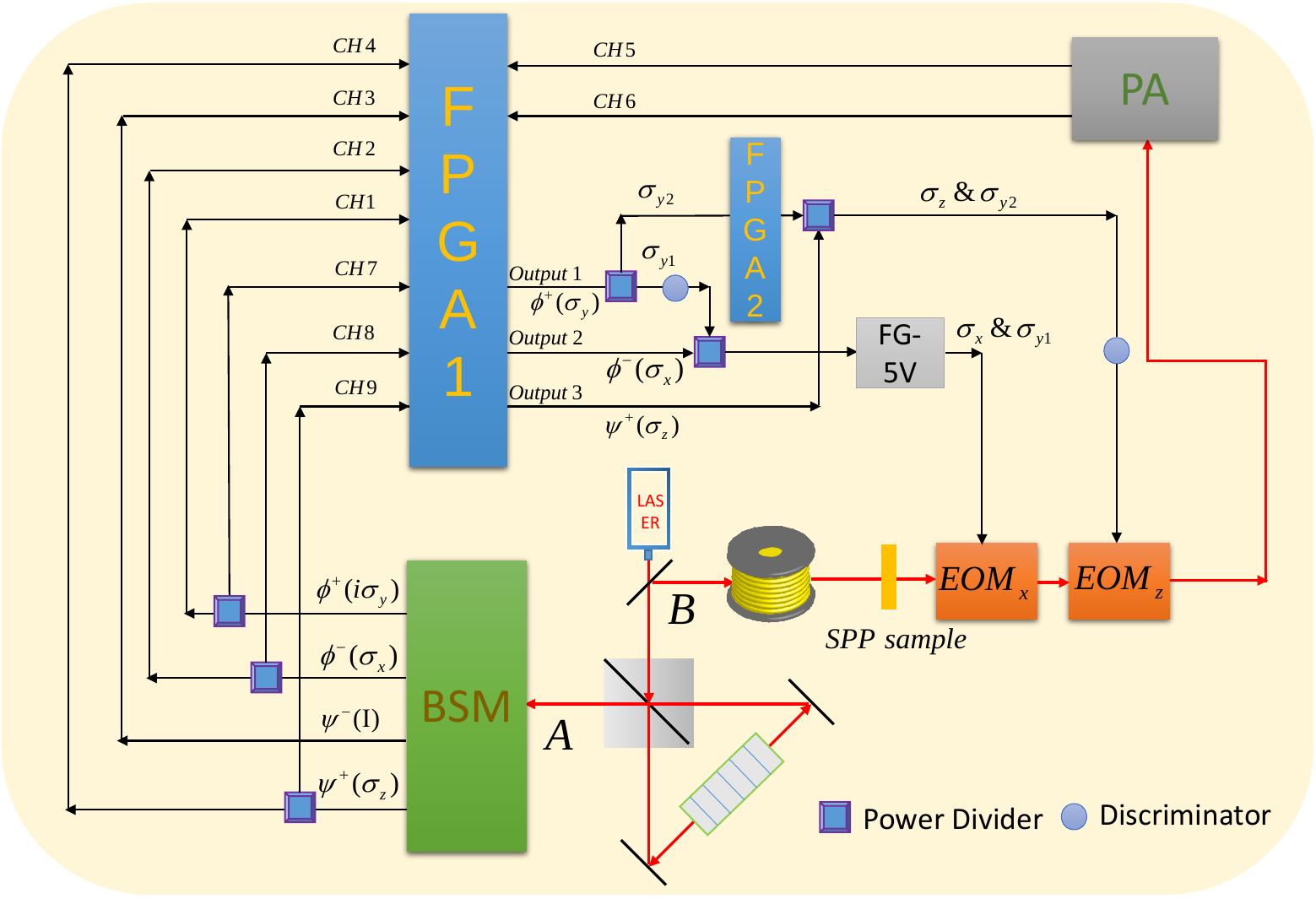}
	\caption{Feed-forward unitary transformation circuit with EOM. The two EOMs perform the Pauli operations acting on photon B to complete the feed-forwrd unitary transformations so as to recover the original quantum states of photon A. The signal of photon A coming out from the BSM is divided into two copies, one is used for coincidence detection with the photon B and another is used for triggering the EOMs. The two FPGAs are exploited to tune the delay of triggering signals acting on the two EOMs. The delay of signal $\vert\Phi^-\rangle^{01}_{A}$ from CH8 to Output2 of FPGA1 is tuned to trigger the EOM$_x$. The delay of signal $\vert\Psi^+\rangle^{01}_{A}$ from CH9 to Output3 is tuned to trigger the EOM$_z$. The signal $\vert\Phi^+\rangle^{01}_{A}$ coming out from Output1 is split into $\sigma_{y1}$ and $\sigma_{y2}$ and will trigger EOM$_x$ and EOM$_z$ simultaneously. Note that the FPGA2 is utilized to compensate the delay cost by the function generator (FG). Here, six power dividers are used to divide and combine the signal from different Bell states. The discriminator is used to provide the voltage of external triggering pulse acting on the EOMs. PA: polarization analyser; BSM: Bell-state measurement; FPGA: Field Programmable Gate Array.}
	\label{SIfig:EOM}
\end{figure}
By performing the BSM, four Bell states will be unambiguously discriminated. Then, we need the results of the BSM to be sent from Alice to Bob, who applies the corresponding unitary transformations. The whole feed-forward setup is shown in Fig.~\ref{SIfig:EOM}. The signals from each BSM outcome are used to trigger the electro-optic modulators (EOMs) to implement corresponding Pauli matrix operations. Two EOMs (EOM$_x$: Leysop RTP-X-4-20; EOM$_z$: ConOptics 360-160-4P-LTA) are used to execute the $\sigma_x$ and $\sigma_z$ operations, respectively. There are total four unitary operations \{$i\sigma_y,\sigma_x,I,\sigma_z$\} for the four Bell states. The EOM$_x$ will be triggered to perform the $\sigma_x$ operation corresponding to the result of $\vert\Phi^-\rangle^{01}_{A}$ and EOM$_z$ performs the $\sigma_z$ operation corresponding to the result of $\vert\Psi^+\rangle^{01}_{A}$. If the outcome of the BSM is $\vert\Phi^+\rangle^{01}_{A}$, the two EOMs will be triggered simultaneously to perform $i\sigma_y$ operation ($\sigma_z\cdot\sigma_x=i\sigma_y$). Here, field programmable gate arrays (FPGAs) are utilized to tune the delay between photon A and photon B. Because the inner interval of FPGA and electronic delay of trigger signal from FPGA to EOM have a minimum limit (247 ns and 114 ns), we use an extra 222 m (time delay of $\sim$1110 ns) SMF to allow the free adjustment of the delay time. In our experiment, three channels (CH7, CH8, CH9) of FPGA1 are used to set the delay between the detection signal of photon A for each Bell port and the trigger signal acting on photon B. Furthermore, the trigger signal from the Output1 set by CH7 is divided into two paths and used to trigger EOM$_x$ and EOM$_z$ simultaneously. For EOM$_x$, we need a function generator (FG) to produce a $>$5 V external triggering voltage. Therefore, we add another FPGA2 in the $\sigma_{y2}$ path to compensate the delay caused by FG. This can put the pulse of the two signals (Output1$\rightarrow$ $\sigma_{y1}$ $\rightarrow$ EOM$_x$, Output1$\rightarrow$ $\sigma_{y2}$ $\rightarrow$ EOM$_z$) in the same time window. Through scanning the time of CH7, CH8 and CH9, the delay of trigger signal corresponding to three Bell states ($\vert\Psi^+\rangle^{01}_{A}$, $\vert\Phi^-\rangle^{01}_{A}$ and $\vert\Phi^+\rangle^{01}_{A}$) can be determined, respectively.

In order to obtain a good contrast, we optimize the half-wave voltage of these two EOMs and set the pulse width of trigger signal to be 200 (100) ns for EOM$_x$ (EOM$_z$). The detailed specifications of these two EOMs are listed in Table~\ref{TabBSM61:EOM}.
\begin{table}[!htbp]
\centering
\caption{Some typical parameters of the two EOMs used in our experiment.}
\label{TabBSM61:EOM}
\begin{threeparttable}
\begin{tabular}{| l | c | c | c | c | c | c |}
 \hline
         & Half-wave voltage & Fast axis's angle  & Pauli matrix & Contrast & Pulse width \\
 \hline
 EOM$_x$ & 1.22 $kV$         & 45$^\circ$         & $\sigma_x$   & 77.5     & 200 $ns$ \\
 \hline
 EOM$_z$ & 70.2 $V$          & 0$^\circ$          & $\sigma_z$   & 29.7     & 100 $ns$ \\     
 \hline
\end{tabular}
\end{threeparttable}
\end{table}

\section{Reconstructed density matrix and teleportation fidelity}\label{sec:seven}
\begin{figure}[!htbp]
\centering
	\includegraphics[width=0.9\textwidth]{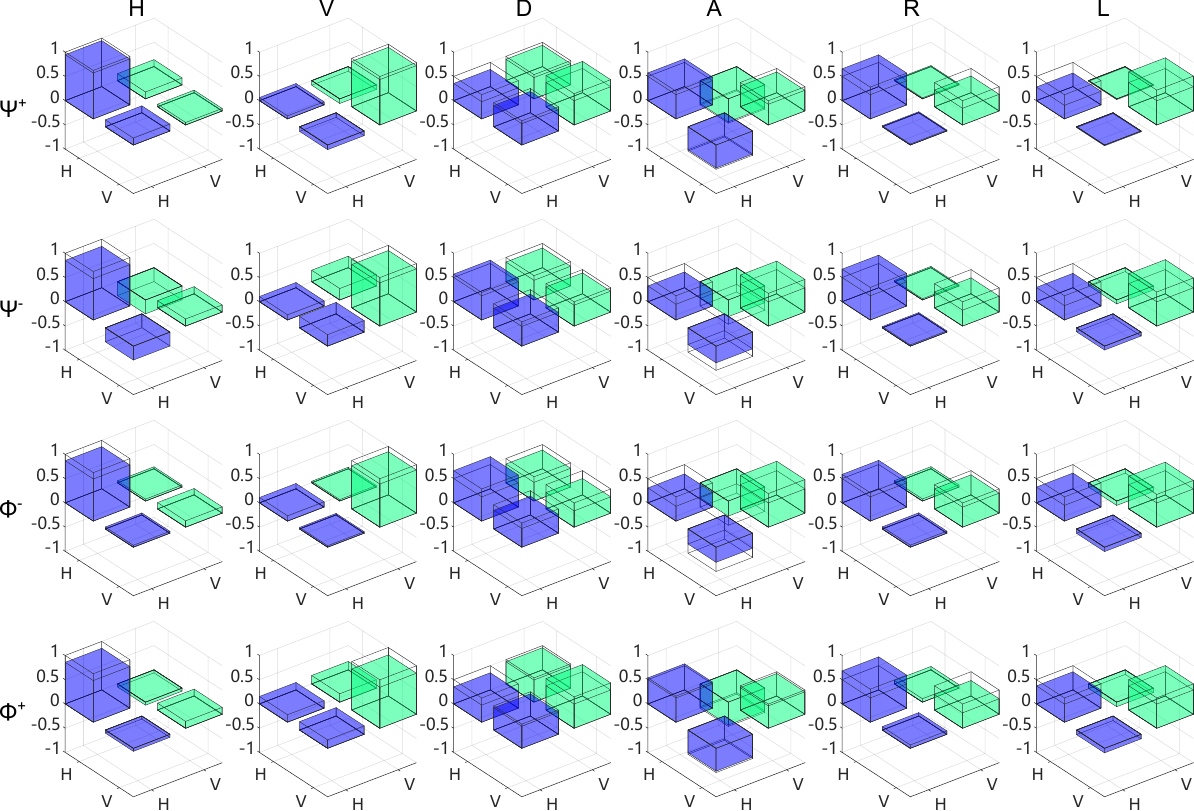}
	\caption{Real parts of the reconstructed density matrices for the six teleported states. The six columns represent the six teleported states and each of the four rows corresponds to one of the possible BSM outcomes. The ideal density matrix $\ket{\phi}_{\text{ideal}}\prescript{}{\text{ideal}}{\bra{\phi}}$ is shown as the wire grid.}
	\label{SI:DMReal}
\end{figure}
\begin{figure}[!htbp]
\centering
	\includegraphics[width=0.9\textwidth]{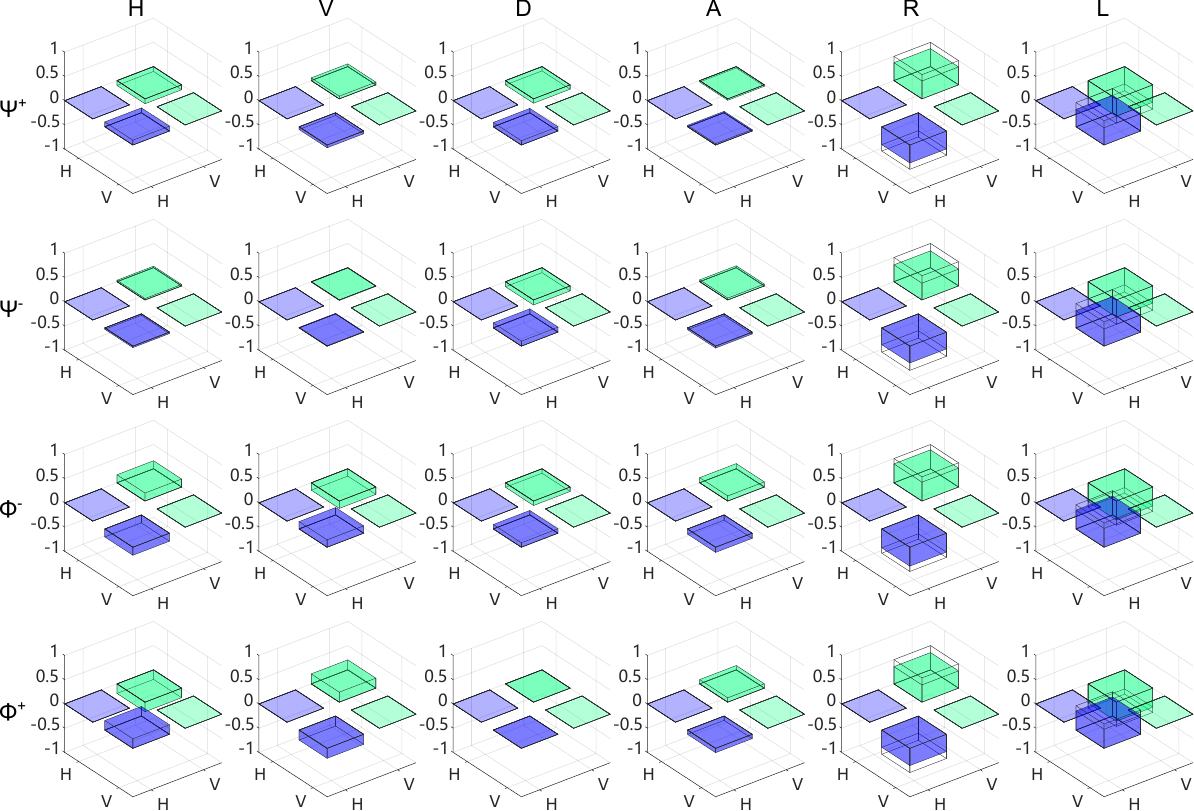}
	\caption{Imaginary parts of the reconstructed density matrices for the six teleported states. Plots are similar to these described in the caption of Fig.~\ref{SI:DMReal}.}
	\label{SI:DMImag}
\end{figure}

We prepare six input states $\ket{H}$, $\ket{V}$, $\ket{D}$, $\ket{A}$, $\ket{R}$ and $\ket{L}$ at the Alice's side. The density matrices for the six teleported quantum states including active feed-forward operations are reconstructed by means of quantum state tomography (QST)~\cite{SJames2001PRA.64.052312}. The real and imaginary parts of the reconstructed density matrices for each of the six states corresponding to the four BSM outcomes are shown in Figs.~\ref{SI:DMReal} and~\ref{SI:DMImag}, respectively. All these figures are for the teleportation with the SPP involved. We can see that the $\ket{H}$ and $\ket{V}$ states have one dominating element. For $\ket{D}$ and $\ket{A}$ states, the four elements have approximate equal weight. The diagonal elements are in opposite sign with the antidiagonal elements for $\ket{A}$ state. For $\ket{R}$ and $\ket{L}$ states, the diagonal elements are real and the antidiagonal elements are imaginary. These features are consistent with the ideal density matrices of these six states.

The teleportation fidelities of these states are calculated with the reconstructed density matrices via $F=\prescript{}{\text{ideal}}{\left\langle\phi\right|}\rho\ket{\phi}_{\text{ideal}}$, where $\ket{\phi}_{\text{ideal}}$ is the ideal quantum state. The uncertainties in state fidelities are calculated using a Monte-Carlo method assuming Poissonian counting statistics. In Table~\ref{Tab2:FidWOSam} and~\ref{Tab3:FidWSam}, we give the obtained fidelity data for both without (Table~\ref{Tab2:FidWOSam}) and with (Table~\ref{Tab3:FidWSam}) the SPP. The experimental measured coincidence counts for the teleportation fidelities are shown in Table~\ref{Tab4:CCWOSam} (without SPP) and~\ref{Tab5:CCWSam} (with SPP). The teleported state $\ket{\phi}$ is projected to detectors D5 and D6 for tomography (see Fig.~1(e) in main text). For the states $\ket{H}$, $\ket{D}$ and $\ket{R}$, $\ket{\phi}$ is projected to detector D5 and the orthogonal state $\ket{\phi}^\perp$ is projected to D6. For $\ket{V}$, $\ket{A}$ and $\ket{L}$ states, $\ket{\phi}$ is projected to D6 and the orthogonal state $\ket{\phi}^\perp$ is projected to D5.
\begin{table}[!htbp]
\centering
\caption{State fidelities without SPP (in units of \%).}
\label{Tab2:FidWOSam}
\setlength{\tabcolsep}{1mm}
{
	\begin{tabular}{c|c|c|c|c|c|c}
	\toprule[1pt]
	            & $\ket{H}$      & $\ket{V}$      & $\ket{D}$      &   $\ket{A}$    &   $\ket{R}$    &    $\ket{L}$  \\
	\midrule[0.2pt]
	$\Psi^+$    & 95.81$\pm$0.27 & 97.22$\pm$0.21 & 95.17$\pm$0.25 & 95.90$\pm$0.25 & 95.20$\pm$0.27 & 97.98$\pm$0.28 \\
	$\Psi^-$    & 88.22$\pm$0.38 & 91.25$\pm$0.36 & 92.71$\pm$0.37 & 88.37$\pm$0.40 & 93.99$\pm$0.26 & 96.85$\pm$0.20 \\
	$\Phi^-$    & 89.47$\pm$0.43 & 91.48$\pm$0.39 & 91.40$\pm$0.34 & 87.50$\pm$0.46 & 92.32$\pm$0.36 & 94.03$\pm$0.30 \\
	$\Phi^+$    & 88.02$\pm$0.40 & 86.46$\pm$0.42 & 95.28$\pm$0.27 & 96.83$\pm$0.21 & 90.18$\pm$0.29 & 92.43$\pm$0.26 \\
	\bottomrule[1pt]
	\end{tabular}
}
\end{table}

\begin{table}[!htbp]
\centering
\caption{State fidelities with SPP (in units of \%).}
\label{Tab3:FidWSam}
\setlength{\tabcolsep}{1mm}
{
	\begin{tabular}{c|c|c|c|c|c|c}
	\toprule[1pt]
	            & $\ket{H}$      & $\ket{V}$      & $\ket{D}$      &   $\ket{A}$    &   $\ket{R}$    &    $\ket{L}$  \\
	\midrule[0.2pt]
	$\Psi^+$    & 94.80$\pm$0.35 & 93.62$\pm$0.36 & 93.97$\pm$0.33 & 97.44$\pm$0.22 & 86.95$\pm$0.32 & 85.75$\pm$0.23 \\
	$\Psi^-$    & 85.32$\pm$0.50 & 89.25$\pm$0.38 & 90.55$\pm$0.42 & 84.50$\pm$0.47 & 84.68$\pm$0.35 & 85.34$\pm$0.28 \\
	$\Phi^-$    & 86.94$\pm$0.55 & 89.04$\pm$0.49 & 88.64$\pm$0.48 & 81.09$\pm$0.73 & 88.98$\pm$0.34 & 88.88$\pm$0.35 \\
	$\Phi^+$    & 85.85$\pm$0.44 & 84.66$\pm$0.38 & 95.67$\pm$0.25 & 97.04$\pm$0.22 & 85.58$\pm$0.39 & 89.31$\pm$0.33 \\
	\bottomrule[1pt]
	\end{tabular}
}
\end{table}

\begin{table}[!htbp]
\centering
\caption{Experimental measured coincidence counts for state fidelities without SPP (total integration time 60s). Each Bell port is made coincidence with the projected state $\ket{\phi}$ and its orthogonal state $\ket{\phi}^\perp$.}
\label{Tab4:CCWOSam}
\setlength{\tabcolsep}{5mm}
{
	\begin{tabular}{ c | c | r | r | r | r | r | r }
	\toprule[1pt]
	\multicolumn{2}{c|}{$\ket{\phi}$} & \multicolumn{1}{c|}{$\ket{H}$} & \multicolumn{1}{c|}{$\ket{V}$} & \multicolumn{1}{c|}{$\ket{D}$} & \multicolumn{1}{c|}{$\ket{A}$} & \multicolumn{1}{c|}{$\ket{R}$} & \multicolumn{1}{c}{$\ket{L}$}  \\
	\hline
	\multirow{2}{*}{$\Psi^+$} & $\ket{\phi}$       & 2,746     & 2,554     & 2,639     & 2,552     & 2,442     & 2,034     \\
	                          \cline{2-8}
	                          & $\ket{\phi}^\perp$ & 120       & 73        & 134       & 109       & 123       & 42        \\
	\hline
	\multirow{2}{*}{$\Psi^-$} & $\ket{\phi}$       & 2,359     & 2,525     & 2,596     & 2,417     & 2,206     & 2,060     \\
	                          \cline{2-8}
	                          & $\ket{\phi}^\perp$ & 315       & 242       & 204       & 318       & 141       & 67        \\
	\hline
	\multirow{2}{*}{$\Phi^-$} & $\ket{\phi}$       & 2,124     & 2,050     & 2,178     & 1,988     & 2,117     & 1,734     \\
	                          \cline{2-8}
	                          & $\ket{\phi}^\perp$ & 250       & 191       & 205       & 284       & 176       & 110       \\
	\hline
	\multirow{2}{*}{$\Phi^+$} & $\ket{\phi}$       & 3,336     & 3,053     & 3,794     & 3,264     & 2,956     & 2,967     \\
	                          \cline{2-8}
	                          & $\ket{\phi}^\perp$ & 454       & 478       & 188       & 107       & 322       & 243       \\
	\hline
	\bottomrule[1pt]
	\end{tabular}
}
\end{table}

\begin{table}[!htbp]
\centering
\caption{Experimental measured coincidence counts for state fidelities with SPP (total integration time 900s). Each Bell port is made coincidence with the projected state $\ket{\phi}$ and its orthogonal state $\ket{\phi}^\perp$.}
\label{Tab5:CCWSam}
\setlength{\tabcolsep}{5mm}
{
	\begin{tabular}{ c | c | r | r | r | r | r | r }
	\toprule[1pt]
	\multicolumn{2}{c|}{$\ket{\phi}$} & \multicolumn{1}{c|}{$\ket{H}$} & \multicolumn{1}{c|}{$\ket{V}$} & \multicolumn{1}{c|}{$\ket{D}$} & \multicolumn{1}{c|}{$\ket{A}$} & \multicolumn{1}{c|}{$\ket{R}$} & \multicolumn{1}{c}{$\ket{L}$}  \\
	\hline
	\multirow{2}{*}{$\Psi^+$} & $\ket{\phi}$       & 2,205     & 2,435     & 2,367     & 1,980     & 1,572     & 1,914     \\
	                          \cline{2-8}
	                          & $\ket{\phi}^\perp$ & 121       & 166       & 152       & 52        & 236       & 318       \\
	\hline
	\multirow{2}{*}{$\Psi^-$} & $\ket{\phi}$       & 1,790     & 2,515     & 2,118     & 2,017     & 1,890     & 1,746     \\
	                          \cline{2-8}
	                          & $\ket{\phi}^\perp$ & 308       & 303       & 221       & 370       & 342       & 300       \\
	\hline
	\multirow{2}{*}{$\Phi^-$} & $\ket{\phi}$       & 1,677     & 2,055     & 1,669     & 1,509     & 1,340     & 1,582     \\
	                          \cline{2-8}
	                          & $\ket{\phi}^\perp$ & 252       & 253       & 214       & 352       & 166       & 198       \\
	\hline
	\multirow{2}{*}{$\Phi^+$} & $\ket{\phi}$       & 2,640     & 3,080     & 3,579     & 2,494     & 2,446     & 2,172     \\
	                          \cline{2-8}
	                          & $\ket{\phi}^\perp$ & 435       & 558       & 162       & 76        & 412       & 260       \\
	\hline
	\bottomrule[1pt]
	\end{tabular}
}
\end{table}

The reduction in state fidelities with SPP compared to that of without (W.O.) SPP can be attributed to different parts of the experiment. The effect of the imperfect optical elements and multiphoton noise can be treated as the white noise and the generated two-photon states are approximated as the Werner states~\cite{SWerner1989PRA.40.4277},
\begin{align}\label{eqSec81:StateNoise}
	\rho_1 = F_{source}\ket{\Psi^-}_{AB}\prescript{}{AB}{\bra{\Psi^-}} +\frac{1-F_{source}}{4}\mathbf{I}
\end{align}
where $\ket{\Psi^-}_{AB} = \frac{1}{\sqrt{2}}\left(\ket{HV}-\ket{VH}\right)$. After the source, photon A enters the BSM and photon B passes through SPP, two EOMs and other optical elements (lens, wave plates, mirror, etc.). We denote the operations as $M=\text{BSM}_A${$\otimes$}(OE$\cdot$EOMz$\cdot$EOMx$\cdot$SPP)$_B$ (OE stands for other optical elements). Finally, the state becomes $\rho_2=F_{tot}M\rho_1M^\dagger+\frac{1-F_{tot}}{4}\mathbf{I}$. The BSM, SPP and two EOMs are responsible for the observed reduction in the measured fidelity and we label their fidelities as $F_{BSM}$, $F_{SPP}$, $F_{EOMx}$ and $F_{EOMz}$. The fidelity reduction caused by remaining optical elements is denoted by $F_{OE}$. Therefore, the state fidelity including all these components can be expressed as:
\begin{align}\label{eqSec82:Fidelity}
	F_{tot} = F_{source}{\cdot}F_{BSM}{\cdot}F_{SPP}{\cdot}F_{EOMx}{\cdot}F_{EOMz}{\cdot}F_{OE}
\end{align}
The non-ideal optical elements (such as PBS, wave plates, mirror and so on) and multiphoton emission reduce the quality of the two-photon entanglement and lead to the 98.34\% fidelity of the source. The imperfect settings of HWP and BD limit the visibility of quantum interference of Bell-state analyser and lead to the 97.87\% fidelity of BSM. To characterize the influences of feed-forward operations on the fidelity, we directly prepare the six states in the SPP setup and measure the state fidelity for individual EOM when moving in and moving out the SPP. By averaging the fidelities over all input states, we obtain the state fidelities of 94.15\% with the SPP and 96.14\% without the SPP for EOMx. For EOMz, the average state fidelities are 95.32\% with the SPP and 97.64\% without the SPP. In addition, we remove the two EOMs and measure the state fidelities both without and with the SPP. This gives the state fidelities of $F_{OE}$ = 98.47\% for W.O. SPP and 95.81\% for with SPP, respectively. Because all the other optical elements are included during the measurement of two EOMs and SPP, we need to eliminate the fidelity $F_{OE}$ and get the net fidelities for these three components. From the above data, we obtain the average state fidelities of $F_{EOMx}$ = 0.9415/0.9847 = 95.61\% (With SPP) and $F_{EOMx}$ = 0.9614/0.9847 = 97.63\% (W.O. SPP) for EOMx. The corresponding state fidelities are $F_{EOMz}$ = 0.9532/0.9847 = 96.80\% (With SPP) and $F_{EOMz}$ = 0.9764/0.9847 = 99.16\% (W.O. SPP) for EOMz. With the SPP involved, the state fidelity is $F_{SPP}$ = 0.9581/0.9847 = 97.30\%. Finally, we calculate the average state fidelities for both W.O. SPP and with SPP to be:
\begin{subequations}\label{eqSec83:AveFid}
	\begin{align}
		F_{W.O.\ SPP} &= F_{source}{\cdot}F_{BSM}{\cdot}F_{EOMx}{\cdot}F_{EOMz}{\cdot}F_{OE} = 91.75\% \\
		F_{With\ SPP} &= F_{source}{\cdot}F_{BSM}{\cdot}F_{SPP}{\cdot}F_{EOMx}{\cdot}F_{EOMz}{\cdot}F_{OE} = 85.35\%
	\end{align}
\end{subequations}
The above quantitative analysis indicates that the excitation of the SPP mode can lead to the deterioration of the beam pattern, which decreases the modulation contrast of the two EOMs and finally results in the reduction of the state fidelity.


\end{document}